\documentclass[11pt]{article}
\usepackage[T1]{fontenc}
\usepackage{lmodern}
\usepackage[letterpaper,margin=1in]{geometry}
\usepackage{microtype}
\usepackage{xcolor}
\usepackage{soul}
\usepackage{amsmath,amssymb,amsthm}
\usepackage{graphicx}
\usepackage{rotating}
\usepackage{enumitem}
\usepackage[sort,nocompress]{cite}
\usepackage[hidelinks]{hyperref}
\usepackage[capitalise,noabbrev]{cleveref}
\crefname{subsection}{section}{sections}

\setlist{nosep,leftmargin=*}

\newcommand{\anon}{\mathbf a}
\newcommand{\cand}{\mathbf b}
\newcommand{\agents}{\mathsf A}

\newtheorem{lemma}{Lemma}
\newtheorem{theorem}{Theorem}
\newtheorem{corollary}{Corollary}

\title{Computation in Anonymous Directed Dynamic Networks: Exact Worst-Case Termination Time and Network-Wise Intractability}
\author{Giuseppe A. Di Luna\thanks{Both authors contributed equally to this research.}
  \\ \normalsize DIAG
  \\ \normalsize Sapienza University of Rome, Italy
  \and Giovanni Viglietta\footnotemark[1]
  \\ \normalsize School of Computer Science and Engineering
  \\ \normalsize University of Aizu, Japan}
\date{}

\begin{document}
\hypersetup{pageanchor=false}
\begin{titlepage}
\maketitle
\thispagestyle{empty}

\begin{abstract}
We determine the exact round complexity of Counting in anonymous directed
dynamic networks.  Agents have no identifiers, but $\ell\geq1$ of them are
designated as leaders.  Every agent knows $\ell$ and
its own leader status, while the population size $n$ is unknown and has to be determined by the agents.
Communication proceeds in synchronous rounds; the directed communication graph
may change arbitrarily and unpredictably from one round to the next, but is assumed to remain strongly connected.  We solve Counting with a distributed algorithm
that makes every agent output $n$ and halt after a number of rounds equal to
$$
 \begin{cases}
  n&\text{if }n=\ell\\
  2n-1&\text{if $n>\ell$ and $\ell=1$}\\
  2n-2&\text{if $n>\ell$ and $\ell\geq2$}.
 \end{cases}
$$
The algorithm works even for networks with loops and parallel arcs and even
when agents learn their outdegree only after sending. We prove that this bound
is worst-case tight over directed multigraphs for every $n$ and $\ell$
with $1\leq\ell\leq n$. Moreover, the local computation is
polynomial (provided that each communication multigraph has at most
$2^{\operatorname{poly}(n)}$ arcs): in every round, each agent solves
at most one linear system with up to $n$ variables. By standard techniques, solving Counting also enables the evaluation of any computable function of the multiset of agents' input values within $2n-1$ communication rounds.

This is the first terminating Counting algorithm for the general
directed network model that uses a polynomial number of communication
rounds, improving the previous $2^{O(n\log n)}$ bound to the exact
linear optimum. This closes the exponential gap for directed
networks and, in undirected networks, removes the quadratic
dependence on the number of leaders in the previous upper bound
and closes the gap between the $3n-2$ upper bound and the
$2n-1$ lower bound for a single leader.
Overall, our results improve earlier Counting bounds from papers appearing at
JACM~2020, FOCS~2022, PODC~2024, SIROCCO~2024, and JACM~2026.

The bound above optimizes the worst case over all networks.
For an individual network, we show that even deciding whether an
agent already has enough local information to determine $n$ is
$\mathsf{coNP}$-complete (provided that the numbers of parallel
arcs are encoded in binary). Thus, worst-case-optimal Counting admits polynomial local computation, whereas exact network-wise optimization is intractable under
standard complexity assumptions.
\end{abstract}

\vfill
{\small\noindent\textbf{Formal verification.}
Formal proofs in Lean of all results stated in this paper are available at
\url{https://github.com/gadiluna/ADDN_lean}.\par}
\end{titlepage}

\pagestyle{empty}
\tableofcontents
\clearpage
\pagestyle{plain}
\setcounter{page}{1}
\hypersetup{pageanchor=true}

\section{Introduction}
Modern distributed systems, including networks of wearable devices, smart devices, and sensors, are characterized by continually changing communication topologies. 
 Understanding the computational foundations of such systems is a central motivation for the
study of \emph{dynamic networks}, in which communication links evolve over
time~\cite{KLO10,CFQS12,APR16}. In this abstraction, algorithms cannot rely on a fixed topology or
wait for the network to become stable.

Another fundamental line of research concerns \emph{anonymous systems}, in
which agents have no unique identifiers~\cite{Angluin80,MCS13}. This abstraction makes it possible to study
coordination without requiring a mechanism for assigning and maintaining
global identities. This property is attractive in highly dynamic environments
and in applications where identifiers raise privacy concerns.
COVID-19 contact-tracing apps illustrate these concerns: even rotating random
identifiers did not eliminate worries about tracking and the disclosure of
personal behavior~\cite{SB20,BaumgartnerEtAl20}.
Anonymous models are also used to study insect colonies and other biological
systems~\cite{GMRL15}.

Nonetheless, anonymity creates the obstacle of symmetry: agents cannot use identities
to distinguish information originating from different participants; this limitation can make basic tasks such as leader election impossible~\cite{Angluin80}.
A standard symmetry-breaking assumption is to designate one agent, or a
known nonzero number of agents, as \emph{leaders}; this enables non-trivial computations~\cite{MCS13,DLV26}.

The \emph{Counting} problem consists of determining the total number \(n\) of agents in the network.
Knowledge of $n$, in turn, enables the evaluation of any computable aggregate function of the agents' input values and the design of termination conditions~\cite{KLO10}. However, terminating Counting poses a fundamental challenge: deciding
whether the information observed so far accounts for the entire network.
An agent may obtain a correct estimate without being able to rule out
a larger, observationally indistinguishable network; the difficulty is
therefore not only to compute $n$ correctly, but to know when the available evidence
excludes every other network size.

\emph{History trees} were introduced to study deterministic computation in anonymous undirected dynamic networks~\cite{DV22}. These combinatorial structures record how agents with identical observation histories become distinguishable as communication proceeds. With a unique leader, this framework yields an algorithm that solves Counting in \(3n-2\) rounds, also enabling the computation of every deterministically computable function within the same number of rounds~\cite{DV22}. Subsequent work extended the framework to multiple leaders, giving polynomial-round terminating algorithms, and obtained linear-round stabilizing algorithms for leaderless Average Consensus~\cite{DLV26}. These results established Counting as a central problem in anonymous dynamic networks.

In this paper, we extend the history-tree approach to obtain
\emph{worst-case-optimal terminating Counting} in anonymous {\bf directed dynamic
networks}. Directionality breaks the reciprocity between sending and receiving, invalidating a basic assumption used by techniques developed in previous work~\cite{DV22,DLV26}.
Besides posing interesting theoretical challenges, directed communication arises in wireless sensor networks with heterogeneous transmission ranges, where one device may receive another's messages without direct communication in the reverse direction~\cite{PS09}.

We consider agents communicating in synchronous
rounds over directed graphs that may change arbitrarily, subject only to
strong connectivity in every round. This is the directed version of the 1-interval connectivity
condition considered in~\cite{KLO10}. 
Every agent knows the number \(\ell\geq1\) of leaders,
but not the network size \(n\), and broadcasts the same message along all
outgoing arcs. We assume only \emph{late outdegree awareness}: an agent
learns how many outgoing arcs it has after sending its message.\footnote{Without any form of outdegree awareness, Counting is impossible in general anonymous directed broadcast networks, even with a unique leader~\cite[Section~3.3]{VIG24}.}
In this setting, the best previously available terminating Counting algorithm
terminates in at most $2^{O(n\log n)}$ rounds, even when the communication graphs
are simple~\cite{VIG24}. 

\subsection{Contributions}
We close the aforementioned exponential gap by determining the exact worst-case termination time
of Counting for every network size and every known nonzero number of
leaders. We give a Counting algorithm that works in all anonymous directed dynamic networks of $n$ agents and has optimal worst-case complexity for every number of leaders $\ell \in [1,n]$ (\Cref{sec:algorithm}). Our algorithm terminates in $n$ rounds when $n=\ell$; in $2n-1$ rounds when $n > \ell$ and $\ell=1$; and in $2n-2$ rounds when $n> \ell$ and $\ell >1$. The algorithm works even in multigraphs, and its local computational complexity is polynomial in $n$, assuming that the network has at most $2^{\operatorname{poly}(n)}$ arcs in every round. Moreover, for every combination of $n\geq 1$ and $\ell \in [1,n]$, we show an exact lower-bound example that matches the upper bound given by our algorithm in the directed multigraph model (\Cref{sec:sharp-bounds}).

By standard history-tree techniques~\cite{DV22,DLV26,VIG24},
our algorithm can be extended to compute the multiset of agents'
input values, and hence any computable function of it, within
\(2n-1\) communication rounds in every case
(\Cref{app:optimizations}).

Furthermore, we consider the local decision underlying earlier termination: whether an agent has enough information to determine $n$. We prove that this test is \(\mathsf{coNP}\)-complete even for networks with a unique leader, provided that the numbers of parallel arcs are encoded in binary (\Cref{sec:network-wise-complexity}).
Thus, worst-case-optimal Counting admits polynomial local computation,
whereas even the basic decision needed for exact network-wise
optimization is intractable under standard complexity assumptions.

Table~\ref{tab:proposed-counting-bounds} compares our results with the best-known results for undirected and directed networks.

Our algorithm closes the gap between the $3n-2$
upper bound and the $2n-1$ lower bound for single-leader terminating
Counting in undirected networks left in~\cite{DV22,VIG24,DLV26}. It also shows that terminating Counting is possible in a linear number of rounds even in anonymous directed dynamic networks, closing the exponential termination gap left in~\cite{VIG24}.

Our $2n-2$ upper bound for $\ell>1$ also applies to undirected
networks, improving the polynomial bounds of~\cite{KM22,DLV26}
to a bound linear in $n$ regardless of $\ell$. In particular,
it removes the quadratic dependence on $\ell$ in the upper bound
of~\cite{DLV26} and resolves the asymptotic dependence on $\ell$
asked about in~\cite{VIG24}. In the directed model, our matching lower bound also strengthens the previous bound of roughly $2n-\ell$ rounds~\cite{DLV26} to $2n-2$.
Thus, somewhat surprisingly, in the directed model with $n>\ell$, a second leader saves
one round, and further leaders give no additional worst-case
improvement.

We remark that, in dynamic networks with unique identifiers, agents can solve
Counting in \(O(n)\) rounds by repeatedly broadcasting all identifiers
they know~\cite{KLO10}. In fact, all agents terminate at round \(n\) if
each halts at the first round \(t\) in which it knows at most \(t\)
identifiers. Our results show that the price of anonymity is at most
\(n-1\) additional rounds, even in directed networks.
A more detailed discussion of Counting in networks with and without
identifiers appears in \Cref{sec:related-work}.

\subsection{Technical Overview}

We use the \emph{history-tree} framework introduced in~\cite{DV22}.
A history tree describes the evolution of a dynamic network through
groups of agents with identical observation histories and the
communication between these groups. Each node represents one such
group; as new observations distinguish agents within it, the node
branches into children representing the resulting groups.
Each agent reconstructs a portion of the history tree from its own
observations; this portion is called its \emph{vista}.

Our algorithm works directly on this tree representation.
Each agent tests candidate network sizes against its vista, using
the recorded communication to compute candidate sizes for the
represented groups and checking them against the way groups split
and the known number of leaders.

The key new ingredient in the proof is a notion of \emph{credit}
assigned to history-tree nodes. Credit provides a common budget
for the rounds examined by the test and the branching in
a vista. We show that an underestimated network size supplies too
little credit to explain both, ruling out premature termination.
Credit is used only in the analysis, not maintained by the agents.

History trees are also instrumental in proving $\mathsf{coNP}$ membership for the problem of deciding whether an agent's local information determines the network size. They can serve as certificates that this information is insufficient, although establishing polynomial bounds on certificate size and verification time requires careful work.

For $\mathsf{coNP}$-hardness, we reduce from the complement of
\textsc{Exact Cover by 3-Sets}~\cite{GJ79}, constructing a network with one leader and
using history trees to describe its vista.
The same vista can arise in a network of a different size exactly
when the instance has an exact cover.

\begin{table}[tbp]
\centering
\footnotesize
\setlength{\tabcolsep}{5pt}
\renewcommand{\arraystretch}{1.25}
\begin{tabular}{@{}p{0.26\linewidth}p{0.16\linewidth}p{0.34\linewidth}p{0.17\linewidth}@{}}
\hline
\textbf{Network model} & \textbf{Leaders} &
\textbf{Termination bound} & \textbf{Reference} \\
\hline
Undirected multigraphs & \(\ell=1\) &
\(3n-2\) & \cite{DV22} \\
Undirected multigraphs & \(2\leq\ell\leq n\) &
\((\ell^2+\ell+1)(n-1)+1\) & \cite{DLV26} \\
Directed simple graphs & \(1\leq\ell\leq n\) &
\(2^{O(n\log n)}\) & \cite{VIG24} \\
\hline
Directed multigraphs & \(\ell=n\) &
\(n\) & This work \\
Directed multigraphs & \(\ell=1<n\) &
\(2n-1\) & This work \\
Directed multigraphs & \(2\leq\ell<n\) &
\(2n-2\) & This work \\
\hline
\end{tabular}
\caption{Deterministic terminating Counting results in anonymous dynamic networks of unknown size \(n\)
and known leader count \(\ell\). The bounds from this work are tight; the others are upper bounds.}
\label{tab:proposed-counting-bounds}
\end{table}

\section{Model and History Trees}
\label{sec:model}

\paragraph{Network.}
A \emph{dynamic network} is a sequence \(\mathcal G=(G_t)_{t\geq1}\) of directed
multigraphs \(G_t=(V,E_t)\) on a fixed set \(V\) of \(n\) \emph{agents}.  Self-loops
and parallel arcs are allowed.  A fixed set \(L\subseteq V\) consists of the
\emph{leaders}; write \(\ell=|L|\geq1\).  For \(A,B\subseteq V\), let \(E_t(A,B)\)
be the multiset of arcs of \(G_t\) from \(A\) to \(B\).  Every \(G_t\) is
strongly connected: for each nonempty
proper set \(X\subset V\),
\[
                 |E_t(X,V\setminus X)|>0.
\tag{SC}
\]

\paragraph{Computation.}
Every agent runs the same deterministic algorithm, receives \(\ell\) as a
common input, and starts in state \(s_0(x)=1\) if it is a leader and
\(s_0(x)=0\) otherwise.
An agent is either \emph{active} or \emph{halted}; initially, all agents
are active.
Communication round \(t\geq1\) uses the graph \(G_t=(V,E_t)\).
Every active agent \(x\) \emph{observes} a multiset \(I_t(x)\)
containing, for every active agent \(y\), one copy of \(s_{t-1}(y)\)
for each arc from \(y\) to \(x\) in \(E_t\).
Equivalently, each active agent sends its current state along every
outgoing arc.
Agent \(x\) learns its outdegree
\(d_t^+(x)=|E_t(\{x\},V)|\) only after this communication.
It then sets its new state to
\(s_t(x)=\mathsf{update}(s_{t-1}(x),I_t(x),d_t^+(x))\)
and computes
\(o_t(x)=\mathsf{out}(s_t(x),\ell)\in\mathbb N\cup\{\bot\}\),
where \(\bot\) means that it produces no output in this round.
If \(o_t(x)\neq\bot\), the agent outputs \(o_t(x)\) and halts;
it neither communicates nor updates its state thereafter.
An algorithm \(\mathcal A=(\mathsf{update},\mathsf{out})\)
solves the \emph{Counting problem} if, for every \(n\geq1\),
every dynamic network on \(n\) agents, and every nonempty leader set
\(L\subseteq V\), all agents eventually output \(n\).
Its \emph{termination time} on a network is the first round by which
every agent has halted.
State size and local computation are unrestricted unless stated otherwise.
When measuring local computation, we assume that an agent is given each distinct received state only once, together with its multiplicity written in binary.

\paragraph{Indistinguishability classes.}
Informally, two agents are \emph{indistinguishable} after \(t\) rounds if no
deterministic algorithm can place them in different states by then.
We define the \emph{indistinguishability classes} by a sequence
\((\mathcal P_t)_{t\geq-1}\) of successively finer partitions of \(V\), where
\(\mathcal P_t\) refines \(\mathcal P_{t-1}\) for every \(t\geq0\).
Initially, \(\mathcal P_{-1}=\{V\}\). The partition \(\mathcal P_0\) groups agents by
leader status: \(\mathcal P_0=\{L,V\setminus L\}\) (or
\(\mathcal P_0=\{L\}\) if \(L=V\)). For \(t\geq1\), agents \(x\)
and \(y\) belong to the same class of \(\mathcal P_t\) precisely when they
belonged to the same class of \(\mathcal P_{t-1}\), have the same
outdegree \(d_t^+(x)=d_t^+(y)\), and, for every class
\(C\in\mathcal P_{t-1}\), have equally many incoming arcs from agents
in \(C\): that is, \(|E_t(C,\{x\})|=|E_t(C,\{y\})|\) (note that these arcs need not come from the same agents).
An induction on \(t\) shows that, for every deterministic algorithm,
agents in the same class of \(\mathcal P_t\) necessarily have the same state
after \(t\) rounds. Thus an observer cannot tell agents in the same class apart
by their messages.

\paragraph{History tree.}
The \emph{history-tree} data structure was introduced in~\cite{DV22} to study
how and when anonymous agents become distinguishable. The history tree of
a network \(\mathcal G\), with leader set \(L\), has \emph{levels}
\(-1,0,1, 2,\ldots\). For each \(t\geq-1\) and
each indistinguishability class \(C\in\mathcal P_t\), level \(t\) contains one abstract node \(v\) representing the agents in \(C\). The \emph{fiber} of \(v\), denoted by
\(\agents(v)\), is the set \(C\). We denote the set of nodes at level \(t\)
by \(\mathcal H_t\). For \(t\geq0\), a node \(v\in\mathcal H_t\) is a
\emph{leader node} if \(\agents(v)\subseteq L\), and a \emph{nonleader node}
otherwise. The \emph{anonymity} of \(v\) is defined as \(\anon(v)=|\agents(v)|\).

\begin{figure}[tbp]
\centering
\includegraphics[width=\linewidth]{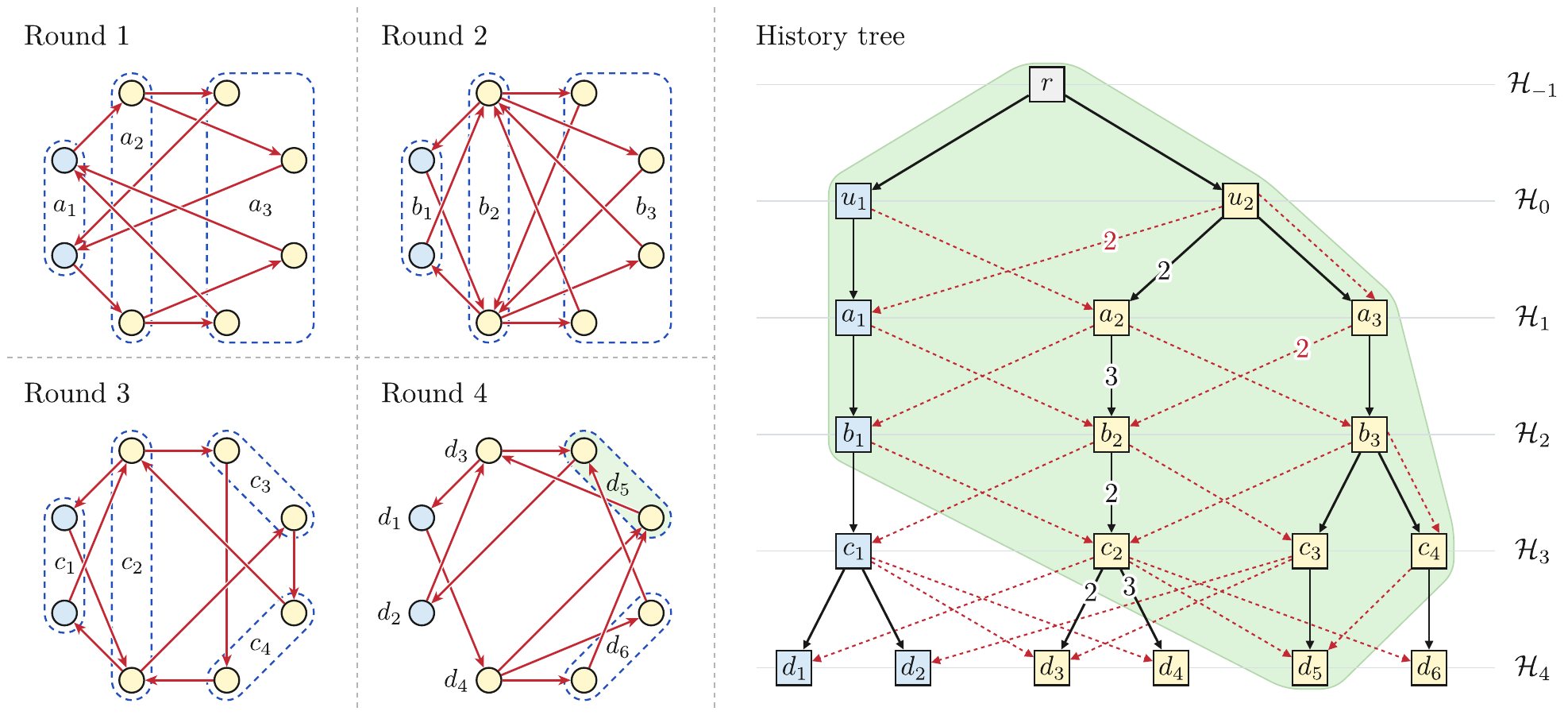}
\caption{Four rounds of a dynamic network and its history tree.
Blue fill denotes leader agents and leader nodes; dashed blue outlines
enclose nonsingleton fibers. Green shading highlights the vista of the
agents represented by \(d_5\). Black-edge labels are outdegrees and
red-edge labels are multiplicities; labels equal to \(1\) are omitted.}
\label{fig:network-history-tree}
\end{figure}

The \emph{parent} of a node \(v\in\mathcal H_t\), with \(t\geq0\),
is the unique node \(u\in\mathcal H_{t-1}\) such that
\(\agents(v)\subseteq\agents(u)\). These parent-child relations form
a rooted tree whose edges are called the \emph{black edges}.
The history tree also contains \emph{red edges}, which represent messages (refer to \Cref{fig:network-history-tree}). Conventionally, both black and red edges point away from the root.
For \(t\geq1\), \(u\in\mathcal H_{t-1}\), and \(v\in\mathcal H_t\),
all agents \(x\in\agents(v)\) have the same number
\(\mu_t(u,v)=|E_t(\agents(u),\{x\})|\) of incoming arcs from agents
in \(\agents(u)\). Whenever this number is positive, the history tree
contains a red edge from \(u\) to \(v\) of multiplicity \(\mu_t(u,v)\).
The agents in \(\agents(v)\) also have a common outdegree
\(d_t(v)=d_t^+(x)\); we place this label on the black edge from \(v\)'s
parent to \(v\).

\paragraph{Vistas.}
Fix an agent \(x\) at round \(t\), and let
\(v\in\mathcal H_t\) be the node whose fiber contains \(x\).  The
\emph{vista} \(\mathcal V_t\) of \(x\) is the subgraph of the history
tree induced by all nodes that lie on a directed path from the root to
\(v\), using black and red edges indifferently.
The node \(v\) itself, which is the vista's unique node at level
\(t\), is called the \emph{bottom} node of \(\mathcal V_t\).  Vistas are considered up to isomorphism
preserving edge types and red multiplicities, the outdegree labels on black
edges, and the leader or nonleader status of every nonroot node.  By construction, two
agents are represented by the same node in \(\mathcal H_t\) exactly when their vistas at
round \(t\) are isomorphic.

As long as all agents are active, each agent can maintain its vista
as its state~\cite{VIG24}. In each round, it match-and-merges its
previous vista with the vistas it observes, then adds the new bottom
node, red edges, and outdegree label.
The stored vista includes the edge labels and the leader or nonleader
status of each nonroot node, but not the fibers or anonymities of its
nodes. Since the root represents all agents, Counting amounts to
determining its anonymity.

\paragraph{Full levels.} For \(-1\leq r\leq t\), let \(\mathcal L_r\) be \emph{level} \(r\)
of \(\mathcal V_t\), namely the set of its nodes that belong to
\(\mathcal H_r\). These nodes are \emph{visible} to \(x\) at level
\(r\), whereas the nodes in \(\mathcal H_r\setminus\mathcal L_r\)
are \emph{hidden}.
We call \(\mathcal L_r\) \emph{full} if
\(\mathcal L_r=\mathcal H_r\), that is, if it contains every
history-tree node at that level.

The \emph{prefix ending at \(\mathcal L_r\)} is the
restriction of \(\mathcal V_t\) to the levels
\(\mathcal L_{-1},\mathcal L_0,\ldots,\mathcal L_r\).  Equivalently, it is
obtained by match-and-merging the round-\(r\) vistas of all agents
represented by nodes in \(\mathcal L_r\).  This prefix is \emph{full} if
\(\mathcal L_r\) is full.  This also makes every earlier level full,
since every node on such a level has a descendant at level \(r\) and every
visible node brings all its ancestors into the vista.

\subsection{Properties of History Trees}
\label{sec:four-properties}

We list four properties of history trees and their vistas, which follow directly from their definitions. As above,
\(\mathcal H_r\) denotes level \(r\) of a history tree, and
\(\mathcal L_r\) denotes level \(r\) of a vista \(\mathcal V_t\).

\begin{enumerate}[label=\textup{H\arabic*.},wide=0pt]
\item \emph{Disjoint refinement.}
      For \(r\geq0\), every \(v\in\mathcal L_r\) has a parent in
      \(\mathcal L_{r-1}\), and the fibers of the nodes in \(\mathcal L_r\) are pairwise
      disjoint. Consequently, for every
      \(u\in\mathcal L_{r-1}\),
      \[
       \sum_{w\text{ child of }u\text{ in }\mathcal L_r}\anon(w)\leq \anon(u)
       \qquad\text{and}\qquad
       \sum_{v\in\mathcal L_r}\anon(v)\leq n.
      \tag{H1}
      \]
      The first inequality is an equality exactly when every child of \(u\) is visible;
      the second is an equality exactly when \(\mathcal L_r\) is full.

\item \emph{Leader accounting.}
      The fiber of every node at level \(r\geq0\) consists entirely of leaders or entirely
      of nonleaders, and every child of that node has
      the same leader status as its parent.
      Therefore,
      \[
       \sum_{v\text{ leader node in }\mathcal L_r}\anon(v)
       \leq\ell,
      \tag{H2}
      \]
      with equality on every full level.

\item \emph{Flow-balance equations.}
      Red multiplicities and outdegree labels count communication arcs exactly:
      for \(r\geq1\), \(u\in\mathcal H_{r-1}\), and
      \(v\in\mathcal H_r\),
      \[
       |E_r(\agents(u),\agents(v))|
          =\mu_r(u,v)\anon(v),
       \qquad
       |E_r(\agents(v),V)|
          =d_r(v)\anon(v).
      \tag{H3}
      \]
      Hence, counting the arcs leaving \(\agents(u)\) by their heads and by their tails gives
      \[
       \sum_{v\in\mathcal H_r}\mu_r(u,v)\anon(v)
       =
       \sum_{w\text{ child of }u\text{ in }\mathcal H_r}
       d_r(w)\anon(w).
      \tag{Flow}
      \label{eq:flow}
      \]

\item \emph{Delay until a level is full.}
      For every vista \(\mathcal V_t\) and \(r\geq0\),
      \[
       t\geq r+n-1 \quad\Longrightarrow\quad
       \mathcal L_r=\mathcal H_r.
      \tag{H4}
      \]
Indeed, trace the vista backward from its bottom node. Parent edges
retain all represented agents; if some agents are still unrepresented,
strong connectivity gives an arc from one of them to a represented
agent. The corresponding red edge makes the sender's node visible
one level earlier. Thus \(n-1\) backward steps suffice to represent
everyone, proving H4. A formal proof appears in
\Cref{lem:tail-exposure} in Appendix~\ref{app:optimizations}
(see also~\cite[Corollary~4.3]{DV22} for undirected networks).
\end{enumerate}

\subsection{Flat Levels}

For \(r\geq0\), a level \(\mathcal L_r\) of a vista is \emph{flat} if every node in
\(\mathcal L_{r-1}\) has exactly one child in \(\mathcal L_r\).

\begin{lemma}
\label{lem:flat-anonymities}
Let \(\mathcal L_r\) be a full and flat level of a vista, with
\(r\geq1\).  The flow equations on \(\mathcal L_r\), together with \(\ell\),
uniquely determine the anonymities of the nodes of \(\mathcal L_r\).\qed
\end{lemma}

An equivalent statement is found in~\cite[Section~3.3]{VIG24}; \Cref{app:flat-anonymities} gives a self-contained proof. Note that the lemma alone does not yield a terminating Counting algorithm. An agent can check whether a level of its vista is flat, but may not know whether it is full and hence whether the lemma applies (see~\cite[Section~5.1]{DLV26} for examples where naive termination strategies fail).

\section{Worst-Case Terminating Counting}
\label{sec:algorithm}

In \Cref{sec:core-algorithm}, we give the core algorithm, which terminates in
\(2n-1\) rounds.\footnote{A simulator for this algorithm is available at \url{https://gadiluna.github.io/ADDN_simulator/}.}  \Cref{sec:sharp-bounds} improves this bound when \(n=\ell\)
or \(\ell>1\), and proves matching lower bounds.

\subsection{The Core Algorithm}
\label{sec:core-algorithm}

\paragraph{Outline.}
Every agent constructs its vista and,
for \(m=\ell,\ell+1,\ldots\), tests candidate \(m\) at round \(2m-1\), using the prefix
ending at \(\mathcal L_m\).  We will prove that every \(m<n\) is rejected,
whereas \(m=n\) is accepted.  The test consists of the following steps.

\begin{enumerate}[label=\arabic*.,wide=0pt]
\item \emph{Check the visible structure.}
      Reject \(m\) unless \(\mathcal L_0\) contains the leader node, each
      \(\mathcal L_r\) with \(0\leq r\leq m\) contains at most \(m\) nodes,
      and every node in \(\mathcal L_r\), for \(0\leq r<m\), has a child in
      \(\mathcal L_{r+1}\).  Reject \(m\) if none of
      \(\mathcal L_1,\ldots,\mathcal L_m\) is
      flat.

\item \emph{Choose the anchor.}
      Let \(h\), with \(1\leq h\leq m\), be the largest index such that
      \(\mathcal L_h\) is flat.  We call \(\mathcal L_h\) the \emph{anchor}.
      Clearly, \(\mathcal L_h\) and \(\mathcal L_{h-1}\) contain the same
      number of nodes.

\item \emph{Solve the anchor.}
      Introduce one unknown \(\cand(v)\) for each node
      \(v\in\mathcal L_h\),
      and solve the flow equations for
      \(u\in\mathcal L_{h-1}\), with both sums restricted to nodes in
      \(\mathcal L_h\), together with
      \[
       \sum_{v\text{ leader node in }\mathcal L_h}\cand(v)
       =\ell.
      \]
      Reject \(m\) unless this system has a unique
      solution over \(\mathbb Q\). Then reject \(m\) unless every \(\cand(v)\) in this solution is a positive integer and \(\sum_{v\in\mathcal L_h}\cand(v)=m\).  We call the resulting values the \emph{candidate anonymities} of the nodes of \(\mathcal L_h\).

\item \emph{Propagate backward.}
      Use disjoint refinement to propagate the candidate anonymities backward,
      and verify that they satisfy the flow-balance equations at every earlier level.
      Namely, for \(r=h-1,h-2,\ldots,-1\) and \(u\in\mathcal L_r\), set
      \[
       \cand(u)=
       \sum_{v\text{ child of }u \text{ in }\mathcal L_{r+1}}\cand(v).
      \]
      Reject \(m\) unless, for every \(1\leq r<h\) and
      \(u\in\mathcal L_{r-1}\),
      \[
       \sum_{v\in\mathcal L_r}\mu_r(u,v)\cand(v)
       =
       \sum_{w\text{ child of }u \text{ in }\mathcal L_r}
       d_r(w)\cand(w).
      \]

\item \emph{Check the suffix.}
      For each \(v\in\mathcal L_h\), let \(q(v)\) be the
      number of its descendants in \(\mathcal L_m\), counting
      \(v\) itself when \(h=m\).  Reject \(m\) if
      \(q(v)>\cand(v)\) for some \(v\).  Otherwise, accept \(m\), output
      \(m\), and halt.
\end{enumerate}

Thus a test solves only one linear system over \(\mathbb Q\).
Everything else consists of counting nodes, following parent-child
relations, and checking the flow equations already recorded in the vista.

\paragraph{Credit.}
To prove the correctness of this algorithm, we assign a \emph{credit} to every node
\(v\) in the prefix ending at the anchor \(\mathcal L_h\).  Intuitively, this
credit is the part of its candidate anonymity that can be matched with actual
agents in its fiber after reserving one agent in that fiber as a representative:
\[
 c(v)=\min\{\anon(v)-1,\cand(v)\}.
\]
The credit is used only in the proof, since the actual anonymities are unknown to the agents.

For \(0\leq r\leq h\), let
\(\Phi_r=\sum_{v\in\mathcal L_r}c(v)\) be the total credit on
\(\mathcal L_r\).  In the following three lemmas and their proofs, we
assume that candidate \(m\) passes Steps~1--4; the level index \(r\)
ranges from \(1\) to \(h\).  Together, the lemmas show that total credit
cannot increase, and that \(m=n\) if two consecutive levels have the
same total credit.

The first lemma compares a node's credit with the total credit of its
children.  It uses only disjoint refinement and backward propagation,
not the flow equations.

\begin{lemma}
\label{lem:credit-refinement-split}
Let \(v_1,\ldots,v_k\) be the children of
\(u\in\mathcal L_{r-1}\) in \(\mathcal L_r\).
Then \(\sum_{i=1}^k c(v_i)\leq c(u)\).
If equality holds and \(\anon(v_j)\leq\cand(v_j)\) for some child $v_j$,
then \(u\) has exactly one child \(v_1\); moreover,
\(\agents(v_1)=\agents(u)\) and \(\cand(v_1)=\cand(u)\).
\end{lemma}

\begin{proof}
Step~1 gives \(k\geq1\), H1 gives
\(\sum_i\anon(v_i)\leq\anon(u)\), and Step~4 gives
\(\sum_i\cand(v_i)=\cand(u)\). Thus,
\[
 \sum_{i=1}^k c(v_i) = \sum_{i=1}^k \min\{\anon(v_i)-1,\cand(v_i)\}
 \leq
 \min\!\left\{\sum_{i=1}^k(\anon(v_i)-1),\sum_{i=1}^k\cand(v_i)\right\}
 \leq \min\{\anon(u)-1,\cand(u)\}=c(u).
\]
Suppose equality holds and \(\anon(v_j)\leq\cand(v_j)\) for some child $v_j$.
Then \(c(v_j)=\anon(v_j)-1<\cand(v_j)\).
Since every other child has credit at most its candidate anonymity,
\(c(u)=\sum_i c(v_i)<\sum_i\cand(v_i)=\cand(u)\).
The definition of credit therefore gives \(c(u)=\anon(u)-1\).
Each visible child \(v_i\) contributes at most \(\anon(v_i)-1\)
units of credit, so \(k\geq2\) would give total credit at most
\(\anon(u)-2<c(u)\), contradicting equality. Thus \(u\) has a unique visible child, $v_1$; if $v_1$ failed to represent an agent in
\(\agents(u)\), its anonymity would be at most
\(\anon(u)-1\), which would again give total credit at most
\(\anon(u)-2<c(u)\). Thus $v_1$ represents all of $\agents(u)$, and Step~4 gives \(\cand(v_1)=\cand(u)\).
\end{proof}

We now use the flow equations to study nodes with the smallest ratio
of actual to candidate anonymity.  For every
\(v\in\mathcal L_{r-1}\cup\mathcal L_r\), define its \emph{ratio} as
\(\rho(v)=\anon(v)/\cand(v)\).  Let
\(\alpha=\min_{v\in\mathcal L_r}\rho(v)\), let
\(\mathcal D=\{v\in\mathcal L_r:\rho(v)=\alpha\}\), and let
\(\mathcal Q\) be the set of parents of nodes in \(\mathcal D\).
The next lemma shows that if the nodes in \(\mathcal D\) collectively
represent the same agents as their parents, then they represent all
agents in the network.

\begin{lemma}
\label{lem:minimum-ratio-closure-split}
If \(\bigcup_{u\in\mathcal Q}\agents(u)
=\bigcup_{v\in\mathcal D}\agents(v)\),
then \(\mathcal L_r\) is full and every node in it has ratio
\(\alpha\).
\end{lemma}

\begin{proof}
Let \(S\) be the common set of agents in the stated equality.
Every child in \(\mathcal H_r\) of a node in \(\mathcal Q\) must
belong to \(\mathcal D\), for any other child would represent agents
on the parents' side of the
equality but not on the children's side.
Counting the arcs leaving \(S\) gives
\begin{equation}
\label{eq:minimum-ratio-flow-split}
 \begin{aligned}
 |E_r(S,V)|
 &=\sum_{v\in\mathcal D}d_r(v)\anon(v)
 =\alpha\sum_{v\in\mathcal D}d_r(v)\cand(v)
 =\alpha\sum_{u\in\mathcal Q}\;\sum_{w\text{ child of }u} d_r(w)\cand(w)\\
 &=\alpha\sum_{u\in\mathcal Q}\;\sum_{w\in\mathcal L_r}
       \mu_r(u,w)\cand(w)
 \leq\sum_{u\in\mathcal Q}\;\sum_{w\in\mathcal L_r}
       \mu_r(u,w)\anon(w)
 \leq |E_r(S,V)|.
 \end{aligned}
\end{equation}
The first equality counts arcs by their tails using H3;
the second uses the ratio \(\alpha\) of each node in \(\mathcal D\).
The third groups the nodes in \(\mathcal D\) by their unique parents
in \(\mathcal Q\); the fourth sums the candidate flow equations
over \(\mathcal Q\) (Step~3 for $r=h$ and Step~4 for $r<h$).
The first inequality uses \(\alpha\cand(w)\leq\anon(w)\).
The last inequality holds because its left-hand side counts only
arcs from \(S\) ending at agents in the fibers of visible nodes.

The first and last expressions of \eqref{eq:minimum-ratio-flow-split} are identical,
so both inequalities hold with equality.
The first forbids arcs from \(S\) to agents in the fiber of a visible
node whose ratio is larger than \(\alpha\); the second forbids arcs
to agents in the fibers of hidden nodes.
Hence every arc from \(S\) ends at an agent represented by a node
in \(\mathcal D\), and therefore at an agent in \(S\).
Since \(S\) is nonempty, strong connectivity forces \(S=V\).
Hence \(\mathcal D=\mathcal H_r\), proving both conclusions.
\end{proof}

We now combine these two facts.  When total credit is preserved,
\Cref{lem:credit-refinement-split} will show that the minimum-ratio
nodes and their parents represent the same agents.
\Cref{lem:minimum-ratio-closure-split} will then make the level
full, and the known leader count will force actual and
candidate anonymities to agree.

\begin{lemma}
\label{lem:credit-loss-split}
For every \(1\leq r\leq h\),
\(\Phi_r\leq\Phi_{r-1}\).
If equality holds for any \(r\), then \(m=n\).
\end{lemma}

\begin{proof}
Summing the inequalities from \Cref{lem:credit-refinement-split}
over \(\mathcal L_{r-1}\) gives \(\Phi_r\leq\Phi_{r-1}\).
Suppose that \(\Phi_r=\Phi_{r-1}\).
Every one of these inequalities must then hold with equality,
so for every \(u\in\mathcal L_{r-1}\),
\begin{equation}
\label{eq:credit-preservation}
 \sum_{v\text{ child of }u\text{ in }\mathcal L_r}c(v)=c(u).
\end{equation}

Let \(\alpha,\mathcal D,\mathcal Q\) be as in
\Cref{lem:minimum-ratio-closure-split}.
Step~3 and backward propagation make the candidate anonymities of
the leader nodes on \(\mathcal L_r\) sum to \(\ell\), whereas H2
bounds their actual anonymities by \(\ell\).
Thus some leader node has ratio at most \(1\), so \(\alpha\leq1\).
Every \(v\in\mathcal D\) therefore satisfies \(\anon(v)\leq\cand(v)\).
Its parent \(u\) satisfies \eqref{eq:credit-preservation}, so
\Cref{lem:credit-refinement-split} implies that \(v\) is the unique child of \(u\) and \(\agents(v)=\agents(u)\).
Thus each parent in \(\mathcal Q\) and its unique child in
\(\mathcal D\) have the same fiber.

\Cref{lem:minimum-ratio-closure-split} now implies that
\(\mathcal L_r\) is full and every node has ratio \(\alpha\).
In particular, its leader nodes have total actual anonymity
\(\ell\) and total candidate anonymity \(\ell\).
Since \(\anon(v)=\alpha\cand(v)\), these two sums give
\(\ell=\alpha\ell\), hence \(\alpha=1\) because \(\ell>0\).
All actual and candidate anonymities on \(\mathcal L_r\) therefore agree.
Step~3 sets the total candidate anonymity at the anchor to \(m\),
and Step~4 preserves this total at every earlier level.
Since \(\mathcal L_r\) is full, its actual anonymities sum
to \(n\). Thus
\(n=\sum_{v\in\mathcal L_r}\anon(v)
=\sum_{v\in\mathcal L_r}\cand(v)=m\).
\end{proof}

The previous lemma bounds the credit that can reach the anchor.
The levels after the anchor give a lower bound:
the anchor must have at least one unit of credit for each
additional visible node.

\begin{theorem}[Core algorithm]
\label{thm:one-leader}
For every \(1\leq\ell\leq n\), the core algorithm makes all agents output \(n\)
and halt at round \(2n-1\). All local computations take time polynomial in $n$ (assuming the network has at most \(2^{\rm{poly}(n)}\) arcs per round).
\end{theorem}

\begin{proof}
Until the first agent halts, every agent maintains its
vista.
Fix a candidate \(m<n\).  If its test reaches Step~5,
\Cref{lem:credit-loss-split}
and integrality show that each of the first \(h\) rounds loses at least
one unit of credit. Step~3 gives total candidate
anonymity \(m\) at the anchor, and backward propagation preserves this total
at every earlier level.
The leader node in \(\mathcal L_0\) has actual and candidate anonymity \(\ell\), and hence
credit \(\ell-1\).  Any remaining credit comes from
the single nonleader node (if it exists and is visible), and is at most the remaining
candidate anonymity \(m-\ell\). Thus
\[
 \Phi_h\leq \Phi_0-h\leq (\ell-1)+(m-\ell)-h = m-h-1.
\]

For each \(v\in\mathcal L_h\), its \(q(v)\) descendants in
\(\mathcal L_m\) have disjoint nonempty fibers.  Hence \(q(v)\leq\anon(v)\), while Step~5 checks that \(q(v)\leq\cand(v)\).  Therefore, observing that $\sum_{v\in\mathcal L_h}q(v)=|\mathcal L_m|$, we have
\[
 \Phi_h = \sum_{v\in\mathcal L_h}c(v)=
 \sum_{v\in\mathcal L_h} \min\{\anon(v)-1,\cand(v)\}
 \geq \sum_{v\in\mathcal L_h}(q(v)-1) 
     =|\mathcal L_m|-|\mathcal L_h|.
\]
Every level after the anchor is nonflat, and Step~1 ensures that
every node before \(\mathcal L_m\) has a child at the next level.
Each of the \(m-h\) remaining transitions therefore increases
the number of visible nodes, giving
\(|\mathcal L_m|-|\mathcal L_h|\geq m-h\).
Combining the two bounds on \(\Phi_h\) yields
\(m-h\leq m-h-1\), a contradiction.
Thus every agent rejects all candidates \(m<n\) and remains active.

Consequently, all agents still maintain their vistas
when \(m=n\) is tested at round \(2n-1\). H4 implies that
\(\mathcal L_n\) is full, and hence the entire prefix ending there is full, as well.  Since
every node before \(\mathcal L_n\) has a child at the next level, the number of nodes cannot increase in all \(n\)
transitions: \(\mathcal L_0\) is nonempty and no level has more than \(n\) nodes.
Thus some level among \(\mathcal L_1,\ldots,\mathcal L_n\) is flat.
\Cref{lem:flat-anonymities} then recovers the actual anonymities at the
anchor.  Disjoint refinement and flow balance make all backward checks
succeed, while disjointness gives
\(q(v)\leq\anon(v)=\cand(v)\) at the anchor.
Hence every agent accepts \(n\) and halts at round \(2n-1\).

Under the stated arc bound, multiplicities, outdegree labels, and
linear-system coefficients have polynomial bit length.
Step~3 solves one linear system with at most \(n\) variables by
exact Gaussian elimination with fractions kept reduced, which takes
time polynomial in \(n\)~\cite[Theorem~4.10]{KV12}. Up to termination, vistas have polynomially many nodes and edges; their construction and all remaining checks also take time polynomial
in \(n\).
\end{proof}

\subsection{Sharp Bounds}
\label{sec:sharp-bounds}

The core algorithm leaves two opportunities to terminate earlier.  We briefly
sketch the sharp upper- and lower-bound arguments here.
\Cref{app:optimizations,app:termination-lower-bounds} contain the complete
technical proofs.

\paragraph{Upper bound: case \(n=\ell\).}
At round \(\ell\), if an agent's vista contains no nonleader node at level
\(\mathcal L_0\), then the agent outputs \(\ell\) and halts. Indeed, if
\(n>\ell\), condition~\textup{(SC)} would make a nonleader visible within
\(\ell\) rounds. Thus, when \(n=\ell\), all agents correctly halt at round \(n\).

\paragraph{Upper bound: case \(\ell>1\).}
Once the previous test has failed,
every agent knows that \(n>\ell\). Assume that \(\ell>1\). Each agent then tests candidates
\(m=\ell+1,\ell+2,\ldots\) using exactly Steps~1--5 of
\Cref{sec:core-algorithm}, except that the test is performed at round
\(2m-2\) (instead of \(2m-1\)), with final level
\(\mathcal L_{m-1}\) (instead of \(\mathcal L_m\)).
Step~1 requires children only through
\(\mathcal L_{m-2}\), Step~2 chooses \(1\leq h\leq m-1\), and Step~5
counts descendants in \(\mathcal L_{m-1}\). The bound on nodes per level
and the total candidate anonymity remain \(m\).
At \(m=n\), H4 makes the prefix ending at \(\mathcal L_{n-1}\)
full. Moreover,
\(\mathcal L_0\) has a leader node and a nonleader node, so
\(\mathcal L_1,\ldots,\mathcal L_{n-1}\) cannot all be nonflat.
Thus the true candidate is accepted at round $2n-2$. Note that this part of the proof did not require $\ell>1$.

It remains to show that the algorithm does not incorrectly terminate when \(\ell<m<n\); for this part, $\ell\geq 2$ is needed. If such a candidate \(m\)
passes with anchor \(\mathcal L_h\), the proof of
\Cref{thm:one-leader} gives
\[
 m-1-h\geq\Phi_h
 \geq|\mathcal L_{m-1}|-|\mathcal L_h|
 \geq m-1-h.
\]
Thus every inequality is tight. This forces the nodes of
\(\mathcal L_{m-1}\) to have singleton fibers, which together
cover all fibers of $\mathcal L_h$. Moreover, at round \(2m-2\), the backward-expansion
argument used for H4 guarantees at least \(m\) represented agents
at level \(\mathcal L_{m-1}\). Since Step~1 permits at most \(m\) nodes in \(\mathcal L_{m-1}\), this level has exactly \(m\) nodes, each representing one agent.

Every visible node from \(\mathcal L_h\) through
\(\mathcal L_{m-1}\) therefore has anonymity equal to its number
of descendants in \(\mathcal L_{m-1}\).
At the anchor, these descendant counts sum to \(m\).
Step~5 bounds each candidate anonymity below by its descendant
count, and the candidate anonymities also sum to \(m\);
hence all candidate anonymities at the anchor are correct.
Let \(S\) be the set of agents represented at \(\mathcal L_h\) (and hence at every subsequent level through \(\mathcal L_{m-1}\)).
Since \(|S|=m<n\), some agents are outside \(S\). This immediately implies that $h=m-1$, for otherwise, in round $h+1$, the agents in $S$ would only receive messages from $S$, contradicting~\textup{(SC)}.

Since $h=m-1$, the leader equation of Step~3 applied to $\mathcal L_h=\mathcal L_{m-1}$ implies that \(S\) contains all \(\ell\) leaders. Now the idea is to count how many nonleaders in \(S\) have become distinguishable from every agent outside \(S\). At round~0, none have: all nonleaders belong to the same fiber, including those outside \(S\). Then, the flow equations for candidate anonymities imply that agents whose whole fiber is already inside \(S\) send messages only to \(S\). Strong connectivity forces a message from these fibers to someone outside their union. That recipient must still be in \(S\), but its old fiber includes outsiders. The recipient receives a message from the sender's fiber, whereas no outsider does. Its new fiber therefore contains no outsiders.

So each round separates at least one additional nonleader in \(S\) from all outsiders. Once separated, it stays separated, because fibers only split.
But there are \(m-1\) rounds and only \(m-\ell\) nonleaders in \(S\). When \(\ell\geq2\), there are too few nonleaders for this to happen in every round. Thus, a false candidate $m<n$ could not have passed. \Cref{app:optimizations} gives the complete argument.

\paragraph{Matching lower bounds.}
The previous combined upper bounds are tight in every case.
For \(n=\ell\), consider two static bidirected paths.  One
consists of \(n\) leaders; the other consists of a chain of \(n\) leaders
followed by one nonleader.  The endpoint leader in each network has the same
vista through round \(n-1\). Hence that leader cannot terminate before round \(n\).

For \(n>\ell\) and \(\ell=1\), two static bidirected paths of
orders \(n\) and \(n+1\), with the leader at one endpoint and a loop at the
other, give the same leader vista through round \(2n-2\).  This construction
and the resulting lower bound already appear in~\cite{DLV26}.

\begin{figure}[tbp]
\centering
\includegraphics[width=\linewidth]{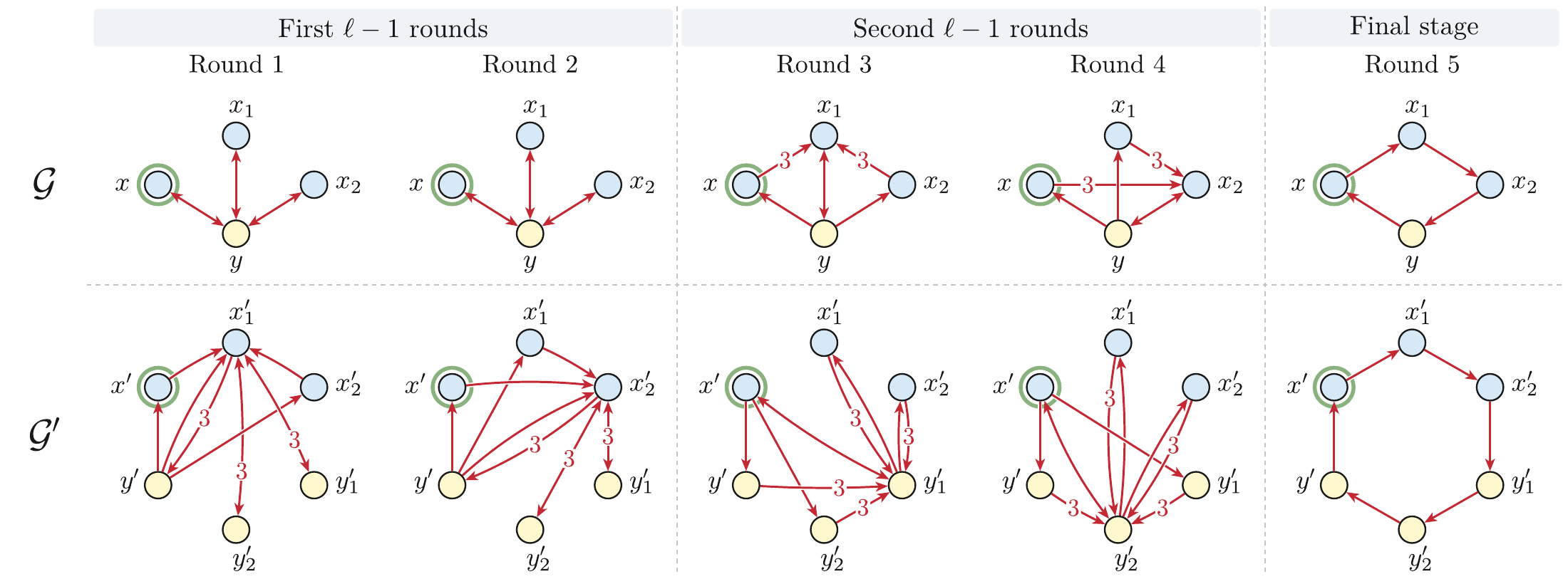}
\caption{The basic lower-bound pair for \(\ell=3\) and \(n=4\).
Blue fill denotes leaders and yellow fill denotes nonleaders.
Green rings mark \(x\) and \(x'\), whose vistas agree through round
\(5=2n-3\). Edge labels are parallel-arc multiplicities; labels equal
to \(1\) are omitted.}
\label{fig:termination-lower-bound}
\end{figure}

Finally, suppose that \(n>\ell\) and \(\ell\geq2\).
We construct two dynamic networks \(\mathcal G\) and \(\mathcal G'\)
of orders \(n\) and \(n+\ell-1\), respectively, each with \(\ell\) leaders.
Let \(x\) and \(x'\) be leaders in \(\mathcal G\) and \(\mathcal G'\),
respectively. We keep their vistas identical through round \(2n-3\).
We first handle \(n=\ell+1\), when \(\mathcal G\) has one nonleader
and \(\mathcal G'\) has \(\ell\) (refer to \Cref{fig:termination-lower-bound}).

\begin{itemize}
\item \emph{First \(\ell-1\) rounds.}
In \(\mathcal G\), every round graph is a bidirected star
centered at its only nonleader.
In each round, \(\mathcal G'\) selects a different leader
other than \(x'\), together with a fixed nonleader.
The selected leader sends to all nonleaders, the selected nonleader
sends to all leaders, and every remaining agent sends only to
the selected leader.
With suitable parallel-arc multiplicities, after each round
all leaders in \(\mathcal G\) and all leaders in \(\mathcal G'\)
not yet selected have the same vista, while all nonleaders
in both networks also have the same vista.

\item \emph{Next \(\ell-1\) rounds.}
In each round, \(\mathcal G\) selects its only nonleader
and a different leader other than \(x\).
Meanwhile, \(\mathcal G'\) selects \(x'\) and a different nonleader.
In each network, the selected leader sends to all nonleaders,
and the selected nonleader sends to all leaders.
Every remaining agent in \(\mathcal G\) sends only to the
selected leader; every remaining agent in \(\mathcal G'\)
sends only to the selected nonleader.

Some previously selected agents may already have different histories,
but their messages reach only the agent newly selected that round.
This agent sends its old state before receiving their information.
Parallel-arc multiplicities match the received multisets and outdegrees
so that, after each round, the leaders in \(\mathcal G\) not yet
selected in this phase have the same vista as \(x'\), while the nonleaders
in \(\mathcal G'\) not yet selected in this phase have the same vista as the only nonleader in \(\mathcal G\).

\item \emph{Final stage.}
Each network becomes a directed cycle through all its agents.
The nonleader in \(\mathcal G\) and the nonleader in \(\mathcal G'\)
left unselected in the second phase send directly to \(x\) and \(x'\),
respectively.
Their identical old states give \(x\) and \(x'\) one further
round of identical vistas.
\end{itemize}
For larger \(n\), add one nonleader to each network at a time.
Initially, route leader information through the new agent before
it reaches the original nonleaders, gaining one extra initial round.
The next two phases still take \(\ell-1\) rounds each,
with their wiring adjusted to preserve the same vista invariants.
Then, in the directed cycle used in the final stage, insert the new agent immediately before \(x\), respectively \(x'\), and repeat this cycle for one additional round:
the extra communication hop keeps the vistas of \(x\) and \(x'\)
identical for one round longer.
Thus adding one nonleader to each network extends the previous construction by two rounds, making vistas agree through round \(2n-3\). Every round graph remains strongly connected and loopless;
\Cref{app:termination-lower-bounds} gives the full construction and proof.

\begin{theorem}
\label{thm:exact-termination}
The exact worst-case termination time for Counting over directed multigraphs is
\[
 T(n,\ell)=
 \begin{cases}
  n&\text{if }n=\ell,\\
  2n-1&\text{if \(n>\ell\) and \(\ell=1\)},\\
  2n-2&\text{if \(n>\ell\) and \(\ell\geq2\)}.
 \end{cases}
\]
All local computations take time polynomial in $n$ (assuming the network has at most \(2^{\rm{poly}(n)}\) arcs per round).\qed
\end{theorem}

\section{Network-Wise Complexity}
\label{sec:network-wise-complexity}
The bound in \Cref{thm:exact-termination} optimizes the worst case
over all networks. However, on a particular network, an agent may have enough
information to determine \(n\) earlier. We now study the complexity
of recognizing when this happens. Before the first agent halts,
every agent can maintain its vista, as described in
\Cref{sec:model}; we ask whether this vista already determines \(n\).
\footnote{This question does not concern whether an agent can stop communicating without preventing others from finishing.}

The input consists of a vista \(\mathcal V\) and the leader count
\(\ell\). It lists the nodes, parents, leader statuses, outdegree
labels, and red multiplicities, with numerical labels in binary;
it does not include fibers or anonymities. A \emph{realization} is
a dynamic network with \(\ell\) leaders in which some agent has
vista \(\mathcal V\). Its \emph{order} is its number of agents.
The \emph{order spectrum} \(\operatorname{Spec}_{\ell}(\mathcal V)\)
is the set of orders of all such realizations. A realizable vista
determines \(n\) exactly when its spectrum is a singleton: two
different compatible orders require different outputs from agents
with the same information.

For a decision problem on arbitrary encodings, define
\textsc{AtMostOneVista} to ask whether
\(|\operatorname{Spec}_{\ell}(\mathcal V)|\leq1\), assigning empty
spectrum to malformed encodings. On vistas that actually occur in
a network, the spectrum is nonempty, so this is precisely the
test for determining \(n\).

\begin{theorem}
\label{thm:main-conp-complete}
\textnormal{\textsc{AtMostOneVista}} is \(\mathsf{coNP}\)-complete.
Hardness holds even for realizable vistas with one leader.\qed
\end{theorem}

We give the main arguments below; \Cref{app:conp-membership,app:conp-hardness}
contain the complete proofs.

\subsection{Membership in \texorpdfstring{\(\mathsf{coNP}\)}{coNP}}
\label{sec:conp-membership}

Membership in \(\mathsf{coNP}\) requires some care. A given vista may be compatible with arbitrarily large networks, so we cannot simply guess two networks of different sizes and describe them explicitly. Describing them by full history-tree prefixes does not immediately solve the problem either, since these prefixes may themselves have exponentially many nodes. We show how to overcome both difficulties and obtain polynomial-size certificates based on history trees, in three steps.

\paragraph{Bounding the network sizes.}
Let $\nu$ be the number of nonroot nodes in the given vista, $t$ its final round,
and $M$ the largest of its numerical labels and the leader count $\ell$.
Suppose two networks of different sizes are compatible with $\mathcal V$.
In each network, choose an agent holding $\mathcal V$ as the
\emph{observer}. An agent is \emph{influential} if its information
can reach the observer by round $t$ through a chain of messages
sent in increasing rounds; the observer itself is influential.
Tracing these dependencies backward shows that at most
$K=(1+\nu M)^t$ agents are influential in either network.

If every agent in both networks is influential, both sizes are
at most $K$ and we are done. Otherwise, consider one of the networks
with non-influential agents, and merge any non-influential nonleaders into a single agent. This preserves strong connectivity and the observer's vista, and yields a network of at most \(K+\ell+1\) agents, some of which are still non-influential.

From such a network, we can obtain a second compatible one by adding one nonleader. In each round, we use this additional agent to subdivide an arc from an influential agent to a non-influential one (which exists by strong connectivity), preserving the sender's outdegree and leaving the vista unchanged. We thus obtain two compatible networks of
consecutive sizes, both at most $K+\ell+2$.

Consequently, whenever two different orders are compatible with \(\mathcal V\), two can be chosen that are at most exponential in the input length (including the binary encoding of \(\ell\)). The binary representations of these orders therefore have polynomial length.

\paragraph{Bounding the history-tree size.}
Even a network of this bounded order might have too many distinct indistinguishability classes. We build a new history tree one level at a time, without changing the network size or the observer's vista. At each new level, we preserve the visible nodes and their fibers. For each node at the previous level, we group the agents in its fiber not represented by visible children into at most one hidden child. If there are hidden agents, we give one of them a singleton fiber, adding at most one extra hidden node at that level. The main difficulty is to preserve strong connectivity without changing visible observations. We achieve this by carefully redirecting communication through that singled-out agent. The resulting prefix through level $t$ has polynomially many nodes, and all its numerical labels have polynomial bit length.

\paragraph{Verifying the certificates.}
Thus, for each of the two chosen orders, we guess a small history-tree prefix through level $t$ matching $\mathcal V$ and $\ell$, including its anonymities and communication multiplicities. We check refinement and anonymity conservation, flow balance, and strong connectivity of a graph describing possible communication between history classes.

These checks can be performed directly on the polynomial-size prefix and do not require constructing the actual network. To prove realizability, we temporarily route messages through an auxiliary vertex for each indistinguishability class. The flow equations ensure that the incoming and outgoing arcs at these vertices can be paired; the main technical step is to choose these pairings so that replacing each pair by a direct arc preserves strong connectivity.

The two history-tree prefixes therefore provide polynomial-size certificates for two different compatible orders, proving membership in $\mathsf{coNP}$. The full proof appears in Appendix E.

\begin{sidewaysfigure}[p]
\centering
\includegraphics[width=\linewidth]{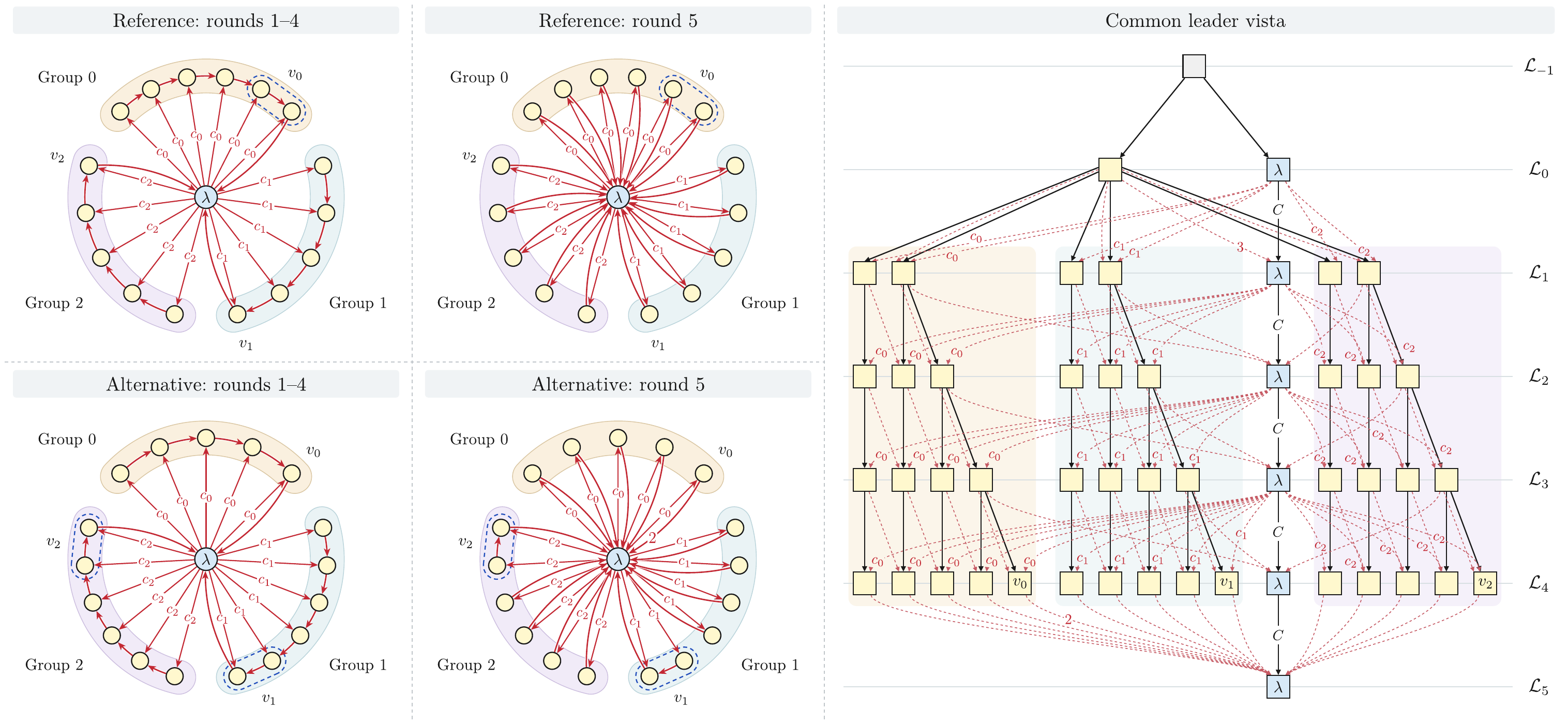}
\caption{The \textsc{X3C} reduction for \(U=\{0,\ldots,5\}\) and
the exact cover \(S_1=\{0,1,2\}\), \(S_2=\{3,4,5\}\).
Here \(q=2\) and \(B=q+3=5\).
Blue fill marks the leader \(\lambda\); yellow fill marks nonleaders.
Shaded sectors identify the special group \(0\) and groups \(1,2\),
corresponding to \(S_1,S_2\).
In rounds \(1\)--\(4\), each group forms a directed path ending at
\(\lambda\). In every round, the leader has \(c_i\) parallel outgoing
arcs to each agent in group \(i\). Each triple is encoded by
\(c_i=B^6+\sum_{e\in S_i}B^e\) for \(i=1,2\), giving
\(c_1=15656\) and \(c_2=19500\); the special coefficient is
\(c_0=2B^6+\sum_{e=0}^{5}B^e=35156=c_1+c_2\).
The reference and alternative group sizes are \((6,5,5)\) and
\((5,6,6)\), respectively. Thus the leader's outdegree is
\(C=6c_0+5c_1+5c_2=5c_0+6c_1+6c_2=386716\)
in both networks, although their orders are \(17\) and \(18\).
The right panel shows their common leader vista after round \(5\).
At level \(4\), \(v_i\) represents the last one or two agents of group
\(i\); a blue dashed outline in a network marks the corresponding
two-agent fiber. The round-\(5\) return arcs give the leader the
same receipts: two copies of the state represented by \(v_0\) and
one from every other level-\(4\) nonleader node.
In the vista, black-edge labels are outdegrees and red-edge labels
are multiplicities; unit labels are omitted throughout.}
\label{fig:conp-reduction}
\end{sidewaysfigure}

\subsection{\texorpdfstring{\(\mathsf{coNP}\)}{coNP}-Hardness}
\label{sec:conp-hardness}

We reduce from the complement of \textsc{Exact Cover by 3-Sets}
(\textsc{X3C})~\cite[Problem~SP2]{GJ79}.
An instance consists of triples \(S_1,\ldots,S_m\) over \(3q\)
elements; an \emph{exact cover} consists of \(q\) triples containing
each element exactly once. We may assume \(q\geq2\), by adding
disjoint triples on fresh elements if necessary.
We construct a one-leader network and output its leader's vista.
This vista will admit a realization of a different order precisely
when the instance has an exact cover (refer to \Cref{fig:conp-reduction}).

\paragraph{The reference network.}
Set \(B=q+3\).
We construct a \emph{reference network} with one leader and \(m+1\) groups of
nonleaders: group \(j\geq 1\) corresponds to triple \(S_j\), and group
\(0\) is an additional special group.
Each triple group contains \(B\) agents, while group \(0\) contains
\(B+1\), giving \(N_0=2+(m+1)B\) agents altogether.
Within each group, arrange the agents along a directed path whose
last agent sends to the leader.
The leader also sends \(c_i\) parallel messages to every agent in
group \(i\). We choose these multiplicities to encode the triples.
Number the universe elements \(0,\ldots,3q-1\), and define
\[
 c_j=B^{3q}+\sum_{e\in S_j}B^e
 \quad(1\leq j\leq m),
 \qquad
 c_0=A=qB^{3q}+\sum_{e=0}^{3q-1}B^e.
\]
In base \(B\), each triple coefficient has three element digits
equal to \(1\), together with a leading digit \(1\).
The target \(A\) has every element digit equal to \(1\) and
leading digit \(q\).
Thus the coefficients of \(q\) selected triples sum to \(A\)
exactly when they form an exact cover: the element digits require
every element to occur once, and carries are prevented due to \(q<B\).

Use the communication graph described above for the first \(B-1\) rounds.
In round \(B\), keep the leader's outgoing arcs, but replace the
path arcs by one arc from every nonleader directly to the leader.
Let \(\mathcal V\) be the leader's vista after this final round.
Thus, the reference network is always a realization of $\mathcal V$, regardless of whether the instance has an exact cover.

\paragraph{The alternative network.}
Suppose the \textsc{X3C} instance has an exact cover.
Remove one agent from group \(0\), and add one agent to each
of the \(q\) triple groups in the cover.
As before, in every round the leader has \(c_i\) parallel
outgoing arcs to every agent in group \(i\).
Because the selected triples cover every element exactly once,
their coefficients sum to \(A\).
The leader therefore loses \(A\) outgoing arcs through the
removal and gains exactly \(A\) through the additions.
Its outdegree is unchanged, although the network now has
\(N_0+q-1>N_0\) agents.

For the first \(B-1\) rounds, arrange each group as a directed
path ending at the leader, as in the reference network.
In round \(B\), replace the path arcs as follows.
The last agent of group \(0\) has two parallel arcs to the leader.
In each selected triple group, put one arc from the second-to-last
agent to the last agent, and one from the last agent to the leader.
Every other nonleader has one arc directly to the leader. \Cref{app:conp-hardness} shows that after \(B\) rounds the leader
has the same vista \(\mathcal V\) in both networks, with \(B\)
visible nodes per group at level \(B-1\).

\paragraph{The converse.}
It remains to prove that if the \textsc{X3C} instance has no exact cover,
every realization of \(\mathcal V\) has exactly \(N_0\) agents.
The \(B\) visible nodes of each group at level \(B-1\) require
at least \(B\) agents, while the leader's prescribed outdegree
permits at most \(q\) additional agents to be represented at
level one.
If level one were not full, strong connectedness would
force the number of represented agents to decrease at each of
the \(B-2=q+1\) transitions to level \(B-1\).
Only \(q\) agents can be lost while retaining \(B\) per group,
so level one must be full.
All outgoing arcs of the leader are then accounted for.
In the absence of an exact cover, the coefficients force the
reference group sizes, and hence order \(N_0\).
\Cref{app:conp-hardness} gives the complete argument.

\section{Related Work}
\label{sec:related-work}

\paragraph{Static networks and local computation.}
In a static anonymous network, an agent's \emph{view} is a
rooted tree recording all information it can learn by exchanging messages.
Tani~\cite{Tani12} shows how to construct and compare compressed views
with polynomial local computation and communication complexity when
the size of a static anonymous network is known. Other work studies
the complexity of recognizing which networks admit a given task.
Yamashita and Kameda~\cite{YK96} prove \(\mathsf{coNP}\)-completeness for recognizing
whether a graph admits election or naming in the port-to-port model;
Chalopin and Paulusma~\cite{CP11} classify these recognition problems
across message-passing and local-computation models. Their input is a
fully specified graph. In our \(\mathsf{coNP}\)-complete problem, the input is an agent's local
history, and the question is whether all compatible dynamic networks
have the same size.

\paragraph{Dynamic networks with IDs.}
Kuhn, Lynch, and Oshman~\cite{KLO10} study exact Counting in
synchronous dynamic networks with unique identifiers and unknown
size. In undirected networks connected in every round, their
algorithms terminate in $O(n)$ rounds using $O(n\log n)$-bit
messages, or in $O(n^2)$ rounds using $O(\log n)$-bit messages.
Both bounds also hold for directed networks that are strongly
connected in every round.
For undirected networks with unrestricted message size, Kuhn,
Moses, and Oshman~\cite{KMO11} improve the stopping criterion
of identifier flooding: agents recognize when they have received
information originating from every node, instead of waiting for
a worst-case time bound. Their criterion is optimal on every
dynamic network and enables Counting in $O(D)$ rounds, where $D$ is
the dynamic diameter, the maximum time needed for information
to travel between any pair of agents.
For strongly connected directed dynamic networks with unique
identifiers and $O(\log n)$-bit messages, Abshoff
et al.~\cite{ABMH13} relate Counting to two-party token
dissemination. They show that transmitting $\Theta(n)$ tokens
between two designated nodes can be reduced to Counting in a
network with $O(n)$ nodes.

\paragraph{Counting with randomization.}
Kuhn, Lynch, and Oshman~\cite{KLO10} also give a randomized
algorithm for approximate Counting in dynamic networks connected
in every round. Their algorithm assumes a known upper bound
$N$ on the network size and a topology chosen independently
of the algorithm's random choices. For any fixed
$0<\varepsilon<1/2$, it terminates in $O(n)$ rounds and returns
an estimate $\widehat n$ satisfying
$|\widehat n-n|\leq\varepsilon n$ with high probability in $N$.
It estimates the size by propagating minima of random values
generated by the agents.
Randomized size estimation has also been studied in peer-to-peer
networks. Jelasity and Montresor~\cite{JM04} use gossip-based
aggregation to estimate global quantities, including network size.
Massouli\'e et al.~\cite{MLKG06} develop estimators based on
random-walk return times and repeated samples of the same peer.
Le Merrer, Kermarrec, and Massouli\'e~\cite{LKM06} compare
random-walk, probabilistic-polling, and gossip-based estimators,
evaluating their accuracy, communication cost, and response to
nodes joining and leaving the network.
These works concern approximate size estimation; here we seek
exact Counting with deterministic termination guarantees.

\paragraph{Anonymous undirected dynamic networks (mass distribution).}
Michail, Chatzigiannakis, and Spirakis~\cite{MCS13} initiated the study
of Counting in anonymous undirected networks whose topology may change
arbitrarily while remaining connected in every round.
Early work by Di Luna et al.~\cite{DBBC14a,DBBC14b} included an
energy-transfer algorithm that lets the leader detect completion
under a degree bound.
Kowalski and Mosteiro~\cite{KM20} obtained the first polynomial-round
algorithm with a unique leader, terminating in \(O(n^5\log^2 n)\).
Their subsequent work~\cite{KM22} extended the approach to a known
number \(\ell\geq1\) of leaders and improved the bound to
\(O(n^{4+\varepsilon}\log^3 n/\ell)\), for every fixed
\(\varepsilon>0\).
These algorithms only work in undirected networks and use mass distribution,
reminiscent of average-consensus techniques~\cite{OT09,NOOT09},
together with tests that detect incorrect estimates of the network size.

\paragraph{Anonymous undirected dynamic networks (history trees).}
Di Luna and Viglietta~\cite{DV22} introduced history trees and obtained
the first linear-round terminating Counting algorithm for anonymous
undirected dynamic networks with a unique leader, with a bound of
\(3n-2\) rounds. A successive work~\cite{DLV26} handles a known
number \(\ell\geq1\) of leaders and, when the network is connected in
every round, gives a termination bound of
\((\ell^2+\ell+1)(n-1)+1\). It also gives a \(2n-2\)-round stabilizing
algorithm for Counting and input-multiset computation, as well as
linear-round algorithms for leaderless Average Consensus. We remark that a stabilizing algorithm need not terminate: agents eventually produce the
correct answer but may not know when it has become correct. 
The same work proves a \(2n-1\) termination lower bound for a unique
leader using static paths with a self-loop~\cite{DLV26}.

\paragraph{Anonymous undirected dynamic networks (communication and memory restrictions).}
In undirected dynamic networks, Di Luna and
Viglietta~\cite{DV25} give polynomial-round algorithms
using \(O(\log n)\)-bit messages, including a
terminating Counting algorithm with a unique leader that takes
\(O(n^3)\) rounds.
Blanc, Di Luna, and Viglietta~\cite{BDV26} study one-bit broadcasts;
with a unique leader, their Counting algorithm stabilizes in
\(O(n^3\log^2 n)\) rounds, and a known upper bound on \(n\) also enables
termination.
Kowalski and Mosteiro~\cite{KM26} study anonymous dynamic networks
with a known nonzero number of leaders, logarithmic message size, and
logarithmic internal working memory. They allow temporary disconnections subject
to a connectivity condition over bounded windows, and give
polynomial-round algorithms for Counting and for
disseminating all inputs with their multiplicities. 

\paragraph{Anonymous directed dynamic networks.}
Charron-Bost and Lambein-Monette~\cite{CBLM24} study
anonymous directed dynamic networks without parallel arcs and with finite
dynamic diameter \(D\): information from any agent can reach every
other agent within \(D\) rounds.
Their method uses a deterministic broadcast
version of the Push-Sum averaging algorithm.
Their algorithm requires \emph{early outdegree
awareness}: each agent knows its current outdegree before sending.
With a known number \(\ell\geq1\) of leaders, their approach can be used
to estimate the fraction \(p=\ell/n\) of leaders. Rounding
\(\ell/\widehat p\), where \(\widehat p\) is the current estimate,
then yields a Counting algorithm whose output eventually stabilizes
at \(n\). Applying the convergence analysis in the proof of
Theorem~5.2 of the full version of~\cite{CBLM24} to this reduction gives a
stabilization bound of \(O(Dn^{2D}\log n)\) rounds. When the network
is strongly connected in every round, \(D\leq n-1\) holds, and so this bound
is \(2^{O(n\log n)}\). The algorithm provides no criterion for
recognizing when the estimate is correct and terminating. 

For directed networks with late outdegree awareness, Viglietta~\cite{VIG24} extends
history trees by incorporating outdegrees and using linear equations
to recover the sizes of indistinguishability classes. This yields
Counting algorithms that stabilize in \(2n-2\) rounds, whereas the
terminating algorithm has a \(2^{O(n\log n)}\) bound for simple round
graphs. Our algorithm closes this gap between stabilization and
termination by attaining the exact worst-case termination bound with
polynomial local computation, including
when loops and parallel arcs are allowed.

\section{Conclusions}
\label{sec:conclusions}

Counting in anonymous directed dynamic networks (allowing loops and
parallel arcs) has an exact worst-case termination bound, attainable
with local computation polynomial in \(n\) if the network has at most
\(2^{\operatorname{poly}(n)}\) arcs in every round.
For a specific network, however, recognizing whether the available
information already determines its size is \(\mathsf{coNP}\)-complete
(for binary-encoded arc multiplicities), even with a unique leader.

One open question is whether the \(2n-2\) termination bound for
Counting remains tight for simple directed round graphs, with no
loops or parallel arcs, when \(2\leq\ell<n\) (for one leader, a \((2-o(1))n\) lower bound
even for static simple undirected networks follows by adapting
the construction of Krebs and Verbitsky~\cite[Section~III]{KV15}).

Another direction concerns \emph{stabilizing Counting}: agents may
revise their outputs and continue communicating, but all outputs
must eventually remain equal to \(n\), without agents having to
detect when this happens. We ask for exact worst-case stabilization
bounds and a corresponding analysis of network-wise optimization
for algorithms that never overestimate \(n\).

Finally, we ask whether \textsc{AtMostOneVista} remains
\(\mathsf{coNP}\)-complete for simple round graphs.
The complexity of \textsc{AtMostOneVista} with unary-encoded
arc multiplicities is also open.

\paragraph{Acknowledgments.}
We thank J\'er\'emie Chalopin and Federico Poloni for discussions concerning a preliminary version of the proof of \Cref{lem:flat-anonymities} using Perron--Frobenius theory.

\paragraph{Disclosure of AI assistance.}
We used the AI assistant GPT-6 Astra to refine and validate the multi-leader
lower-bound construction (\Cref{sec:sharp-bounds,app:termination-lower-bounds})
and to produce formal proofs in the Lean proof assistant for all
results in this paper. The Lean code is available at
\url{https://github.com/gadiluna/ADDN_lean}.

\appendix

\section{Solving a Flat Level}
\label{app:flat-anonymities}

We will now prove \Cref{lem:flat-anonymities} on flat levels. In the terminology of~\cite{VIG24}, the level preceding a flat level is called
\emph{non-branching}, and an equivalent statement to
\Cref{lem:flat-anonymities}
is proved there in Section~3.3.  A closely related statement appears in
Section~4.2 of the full version of~\cite{CBLM24}.
In the following, we prove it within our framework; notably, our proof does not use Perron--Frobenius theory and is self-contained.

\begingroup
\edef\savedflatlevellemmacounter{\number\value{lemma}}
\setcounter{lemma}{0}
\begin{NoHyper}
\begin{lemma}
Let \(\mathcal L_r\) be a full and flat level of a vista, with
\(r\geq1\).  The flow equations on \(\mathcal L_r\), together with \(\ell\),
uniquely determine the anonymities of the nodes of \(\mathcal L_r\).
\end{lemma}
\end{NoHyper}
\setcounter{lemma}{\savedflatlevellemmacounter}
\endgroup

\begin{proof}
Because \(\mathcal L_r\) is flat, every \(u\in\mathcal L_{r-1}\)
has a unique child \(v\in\mathcal L_r\); since \(\mathcal L_r\) is full,
\(\agents(u)=\agents(v)\).
Let \(k=|\mathcal L_r|\), and index these parent-child pairs as
\((u_i,v_i)\), for \(1\leq i\leq k\).

Form the square matrix \(M\) whose row \(i\) corresponds to \(u_i\)
and whose column \(j\) corresponds to \(v_j\), with entries
\[
 M_{ij}=
 \begin{cases}
  \mu_r(u_i,v_j)-d_r(v_j)&\text{if \(i=j\)},\\
  \mu_r(u_i,v_j)&\text{otherwise}.
 \end{cases}
\]
Let \(x\) be the vector of anonymities, with
\(x_i=\anon(v_i)>0\) for \(1\leq i\leq k\).
Since \(\mathcal L_r\) is full, the flow equation
\textup{(Flow)} gives \(Mx=0\), with \(x>0\).

For \(i\ne j\), the entry \(M_{ij}\) is nonnegative and is positive
exactly when \(G_r\) has an arc from \(\agents(v_i)\) to
\(\agents(v_j)\). Thus the digraph with arcs \(i\to j\) whenever
\(i\ne j\) and \(M_{ij}>0\) is obtained from \(G_r\) by contracting
these fibers and discarding loops. It is therefore strongly connected.

We prove that every solution \(Mz=0\) is proportional to \(x\).
Put \(\lambda=\max_i z_i/x_i\) and \(w=\lambda x-z\).
Then \(w\geq0\), \(Mw=0\), and some coordinate \(w_i\) is zero.
The equation in row \(i\) gives
\[
 \sum_{j\ne i}M_{ij}w_j=0.
\]
Every summand is nonnegative, so \(w_j=0\) for every \(j\ne i\)
with \(M_{ij}>0\).
Repeating this argument along directed paths shows that every
coordinate is zero. Hence \(z=\lambda x\), and the nullspace of
\(M\) is spanned by \(x\).

Finally, since \(\mathcal L_r\) is full, H2 gives
\[
 \sum_{v\text{ leader node in }\mathcal L_r}\anon(v)=\ell.
\]
Since \(z=\lambda x\), the entries of \(z\) corresponding to leader
nodes sum to \(\lambda\ell\). If \(z\) also satisfies the leader-count
equation above, this sum equals \(\ell\). Since \(\ell>0\), we obtain
\(\lambda=1\) and hence \(z=x\). Thus the flow equations together
with the leader-count equation uniquely determine the anonymities.
\end{proof}

\section{Causal Support}
\label{app:causal-support}

Fix an agent \(x\) at round \(t\).  Its \emph{causal support} is the
sequence \(\Gamma_0,\ldots,\Gamma_t\) defined by
\[
 \Gamma_t=\{x\},\qquad
 \Gamma_{r-1}=\Gamma_r\cup
 \{y\in V:|E_r(\{y\},\Gamma_r)|>0\}
 \quad(1\leq r\leq t).
\]
Thus \(\Gamma_r\) consists of the agents whose round-\(r\) states are
used, directly or through later states, to construct the round-\(t\)
vista of \(x\). For \(0\leq r\leq t\), a node
\(v\in\mathcal H_r\) belongs to this vista precisely when
\[
                  \agents(v)\cap\Gamma_r\neq\varnothing.
\]
This follows by backward induction from the bottom node: a preceding node
enters the vista as a parent or as the tail of a red edge, corresponding
to the two terms in the recurrence for \(\Gamma_{r-1}\).

\section{Sharp Upper Bounds}
\label{app:optimizations}

We prove the two improvements summarized in \Cref{sec:sharp-bounds}.
The first detects that every agent is a leader. For the second,
we count the nonleader fibers contained in a fixed proper set of
agents to show why two or more leaders permit termination one round
earlier.

\begin{lemma}
\label{lem:tail-exposure}
In a vista \(\mathcal V_t\), the fibers of the nodes in
\(\mathcal L_r\), where \(0\leq r\leq t\), together contain at
least \(\min\{n,t-r+1\}\) agents.
\end{lemma}

\begin{proof}
Use the causal support from \mbox{\Cref{app:causal-support}}, starting with
\mbox{\(\Gamma_t=\{x\}\)}. Whenever \mbox{\(\Gamma_j\neq V\)}, strong connectivity
gives an arc from its complement into it in round \mbox{\(j\)}, so
\mbox{\(|\Gamma_{j-1}|>|\Gamma_j|\)}. Hence
\mbox{\(|\Gamma_r|\geq\min\{n,t-r+1\}\)}, and every agent of \mbox{\(\Gamma_r\)}
belongs to a visible fiber.
\end{proof}

\paragraph{Case \(n=\ell\).}
At round \(\ell\), if an agent's level \(\mathcal L_0\) contains
no nonleader node, it outputs \(\ell\) and halts. If \(n>\ell\),
\Cref{lem:tail-exposure} places at least \(\ell+1\) agents in the
visible level-zero fibers, so a nonleader node is visible.
Thus the test succeeds exactly when \(n=\ell\), and all agents
then halt at round \(n\).

To analyze the multiple-leader case, we use the following fact about
the full history tree. If a proper set contains all leaders and
every fiber contained in that set sends only within it, strong
connectivity forces another nonleader fiber to become contained
in the set at each round.

\begin{lemma}
\label{lem:forward-fibers}
Let \(S\subsetneq V\) contain every leader. Suppose that, for each
\(1\leq r\leq h\), every node in \(\mathcal H_{r-1}\) whose fiber
is contained in \(S\) has no outgoing arc from that fiber to
\(V\setminus S\) in round \(r\). Then \(h\leq |S|-\ell\).
\end{lemma}

\begin{proof}
Count the nonleader fibers contained in \(S\). There are none at
level zero, because the sole nonleader fiber contains \(V\setminus S\).
Every child fiber of a counted fiber is still contained in \(S\),
so this count cannot decrease.

In round \(r\), let \(T\) be the union of the fibers at level
\(r-1\) contained in \(S\). It contains every leader and is a
nonempty proper subset of \(V\). Strong connectivity gives an
arc from an agent \(a\in T\) to an agent \(b\notin T\).
The hypothesis forces \(b\in S\setminus T\).
Let \(A\subseteq T\) be the old fiber containing \(a\).
The agent \(b\) receives an arc from \(A\), whereas no agent
outside \(S\) does. The definition of indistinguishability therefore
places \(b\)'s new fiber wholly inside \(S\), although its old
fiber was not contained in \(S\).
This is a new nonleader fiber, distinct from all children of the
previously counted fibers; it cannot be a leader fiber because all
old leader fibers were already contained in \(T\).

The count increases by at least one per round and is at most
\(|S|-\ell\), proving the bound.
\end{proof}

\paragraph{Case \(\ell>1\).}
Assume that the all-leader test has failed and \(\ell\geq2\).
At round \(2m-2\), test candidate \(m>\ell\) by Steps~1--5 of
\Cref{sec:core-algorithm}, with final level \(\mathcal L_{m-1}\).
Thus Step~1 requires children only through \(\mathcal L_{m-2}\),
Step~2 chooses \(1\leq h\leq m-1\), and Step~5 counts descendants
in \(\mathcal L_{m-1}\), including the node itself when \(h=m-1\).
The bound on nodes per level and the total candidate anonymity
remain \(m\). The credit lemmas apply unchanged, since the test
checks the same refinement and flow equations up to the anchor.

\begin{lemma}
\label{lem:multi-leader-test}
For \(\ell\geq2\), this test rejects every candidate
\(\ell<m<n\) and accepts \(m=n>\ell\) at round \(2n-2\).
\end{lemma}

\begin{proof}
Suppose \(m<n\) passes with anchor \(\mathcal L_h\).
The proof of \Cref{thm:one-leader} gives
\begin{equation}
 m-1-h\geq\Phi_h
 \geq|\mathcal L_{m-1}|-|\mathcal L_h|
 \geq m-1-h.
 \label{eq:short-audit-credit}
\end{equation}
Every inequality is tight. For each anchor node \(v\), let \(q(v)\)
count its descendants in \(\mathcal L_{m-1}\).
Disjoint refinement and Step~5 give
\(\anon(v),\cand(v)\geq q(v)\). Equality in
\eqref{eq:short-audit-credit} gives \(c(v)=q(v)-1\), so
\(\anon(v)=q(v)\). Its \(q(v)\) descendants have disjoint nonempty
fibers contained in \(\agents(v)\); they must therefore have
singleton fibers whose union is \(\agents(v)\).
By \Cref{lem:tail-exposure} and the total candidate anonymity,
\[
 m\leq|\mathcal L_{m-1}|=\sum_{v\in\mathcal L_h}q(v)
 \leq\sum_{v\in\mathcal L_h}\cand(v)=m.
\]
Thus \(\anon(v)=\cand(v)=q(v)\) at the anchor. Its fibers form
a set \(S\) of exactly \(m\) agents, also represented by the
singleton fibers at level \(m-1\). The leader equation shows
that \(S\) contains all \(\ell\) leaders.

If \(h<m-1\), the union of visible fibers is \(S\) at level
\(h+1\) as well: it lies between the identical unions at levels
\(h\) and \(m-1\). Strong connectivity gives an arc from
\(V\setminus S\) into \(S\) in round \(h+1\), making a predecessor
outside \(S\) visible at level \(h\), a contradiction.
Hence \(h=m-1\).

Every earlier node whose fiber meets \(S\) is visible, because
each agent of \(S\) has a visible descendant at level \(m-1\).
Backward propagation from these singleton fibers gives
\(\cand(v)=|\agents(v)\cap S|\) for every
\(v\in\mathcal L_r\), \(0\leq r\leq m-1\).
Fix \(1\leq r\leq m-1\) and \(u\in\mathcal H_{r-1}\) whose
fiber \(A=\agents(u)\) is contained in \(S\).
The node \(u\) and all its children are visible, so the checked
flow equation gives
\[
 |E_r(A,V)|
 =\sum_{v\text{ child of }u\text{ in }\mathcal L_r}
      d_r(v)\cand(v)
 =\sum_{v\in\mathcal L_r}\mu_r(u,v)\cand(v)
 =|E_r(A,S)|.
\]
Thus every old fiber contained in \(S\) sends only to \(S\)
in each of the first \(m-1\) rounds.
\Cref{lem:forward-fibers} now gives
\(m-1\leq |S|-\ell=m-\ell\), contradicting \(\ell\geq2\).

Finally, when \(m=n>\ell\), H4 makes the prefix ending at
\(\mathcal L_{n-1}\) full at round \(2n-2\).
Level \(\mathcal L_0\) has one leader node and one nonleader node.
If the next \(n-1\) levels were all nonflat, the last would have
at least \(n+1\) nodes. Hence a flat level exists;
\Cref{lem:flat-anonymities} recovers its actual anonymities,
and every remaining check succeeds.
\end{proof}

Perform the all-leader test first. After it fails, use the appropriate test, depending on whether \(\ell=1\) or \(\ell\geq2\),
testing \(m=\ell+1,\ell+2,\ldots\).
Every false candidate is rejected at every agent, so all agents
remain active until they halt together at the round stated in
\Cref{thm:exact-termination}.

\paragraph{Input multisets.}
Assume that each agent is given an input value and must evaluate
a fixed computable function of the multiset of all inputs. Neither
\(n\) nor the other agents' input values are initially given.
Following~\cite{DV22,DLV26,VIG24}, each agent labels its initial
bottom node and every new bottom node with its input value.
Match-and-merge preserves these labels and leader status, so the
agents in each nonroot node's fiber have the same input.
Perform the core algorithm's candidate tests, interpreting Step~1
as requiring at least one visible leader node at level zero.
For a candidate \(m\) passing Steps~1--4, the visible level-zero
leader nodes together have credit at most \(\ell-1\): their fibers
contain at most \(\ell\) agents, and each node contributes at most
its anonymity minus one. Nonleader nodes contribute at most the
remaining candidate anonymity \(m-\ell\), so \(\Phi_0\leq m-1\).
The rest of the correctness proof of \Cref{thm:one-leader} applies
unchanged. Thus \(n\) is accepted at round \(2n-1\), and the anchor
anonymities are correct. Summing these anonymities over nodes with
each input value gives its multiplicity. Each agent then evaluates
the function, outputs its value, and halts.

\section{Termination Lower Bounds}
\label{app:termination-lower-bounds}

The lower bounds compare designated leaders in dynamic networks of different
orders but with the same number \(\ell\) of leaders. Equal round-\(t\)
vistas therefore give equal states after \(t\) rounds, so the
two leaders make the same terminal decision.  They cannot both terminate
correctly by that round.

\paragraph{Case \(n=\ell\).}
Let the smaller network be the static bidirected path
\(x_0,x_1,\ldots,x_{n-1}\), with every agent a leader.  The larger network
is the static bidirected path \(x_0,x_1,\ldots,x_{n-1},y\), where the
\(x_i\)'s are leaders and \(y\) is a nonleader. Both networks have
\(\ell=n\) leaders. Observe \(x_0\) in both networks.

The states of the two copies of \(x_i\) agree through every round \(t\) with
\(i+t\leq n-1\).  This follows by induction on \(t\).  Initially the two
agents have the same leader status.  In the induction step, the relevant
copies have the same outdegree and receive the same multiset of states from
corresponding neighbors.  The different degree of \(x_{n-1}\) is learned
only after the first communication round.  Taking \(i=0\) shows that the two
endpoint leaders have the same vista through round \(n-1\). Therefore the
smaller network cannot terminate before round \(n\). This
also covers \(n=1\), when the initial vistas are identical.

\paragraph{Case \(\ell=1\).}
For \(m\geq2\), let \(P_m^\circ\) be the static bidirected path
\(x_0,x_1,\ldots,x_{m-1}\).  Its unique leader is \(x_0\), and
\(x_{m-1}\) has one self-loop.
{Run the vista-construction algorithm on \(P_m^\circ\) and
\(P_{m+1}^\circ\).  At round \(t\), all agents farther than \(t\) edges from
their leader have the same vista: initially they are nonleaders, and at each
step they have degree two and receive two copies of the preceding common
vista.  Corresponding agents therefore agree through round \(m-1\): until
then, the loop at the shorter endpoint returns the same vista as the
continuing edge in the longer path.  A difference can first arise there at
round \(m\), then needs another \(m-1\) rounds to reach the leader.  Thus
the leader vistas agree through round \(2m-2\), and termination requires
\(2m-1\) rounds.}  This construction and bound already appear
in~\cite{DLV26}.

\paragraph{Case \(\ell>1\).}
We first describe a strongly connected graph in which every agent except
one receives information from a single sender.

\begin{lemma}\label{lem:two-center-graph}
Let \(U,W\) be disjoint nonempty sets of agents, with \(|U|=\ell\).
Choose \(z\in U\), \(z'\in W\), and \(c\in\{1,\ell\}\).
An arrow \(X\xrightarrow{d}Y\) denotes \(d\) arcs from each agent of \(X\)
to each agent of \(Y\); an agent may replace a singleton set.
The graph on \(U\cup W\) consisting of exactly the arcs
\[
 z\xrightarrow{c}W,\qquad
 z'\xrightarrow{1}U,\qquad
 U\setminus\{z\}\xrightarrow{\ell/c}z,\qquad
 W\setminus\{z'\}\xrightarrow{\ell}z
\]
is loopless and strongly connected.
Each agent of \(U\setminus\{z\}\) receives only \(z'\)'s old vista and has
outdegree \(\ell/c\); each agent of \(W\) receives \(c\) copies of \(z\)'s
old vista and has outdegree \(\ell\).
\end{lemma}

\begin{proof}
Every agent other than \(z\) reaches \(z\) directly. Conversely, \(z\)
reaches \(W\) directly and \(U\) through \(z'\), proving strong connectivity.
The sets are disjoint and the return bundles exclude \(z\) as a sender,
so there are no loops. The listed arcs give exactly the stated incoming
vistas and outdegrees.
\end{proof}

The construction uses an agent's matching history for one last round:
its current incoming information and outdegree cannot change the state it
has already sent. Two phases apply this observation to the leaders and
nonleaders in turn. A path of added nonleaders then supplies one initial
and one final round of delay per added agent.

\begin{theorem}
\label{thm:several-leader-vista}
For every \(n>\ell\ge2\), there are dynamic networks of orders \(n\) and
\(n+\ell-1\), each with exactly \(\ell\) leaders, whose designated leaders
have identical vistas through round \(2n-3\).
Every round graph is a strongly connected loopless directed multigraph,
with arc multiplicities at most \(\ell\) and outdegrees at most \(\ell^2\).
\end{theorem}

\begin{proof}
We first describe two phases for \(n=\ell+1\), using networks
with \(\ell+1\) and \(2\ell\) agents. We call these networks the
\emph{cores}: for larger \(n\), we add nonleaders to each and
adjust the communication arcs.
The smaller core has leaders \(x_1,\ldots,x_\ell=x\) and one
nonleader \(y\); the larger has leaders
\(x'_1,\ldots,x'_\ell=x'\) and nonleaders
\(y'_1,\ldots,y'_\ell=y'\). We compare the vistas of \(x,x'\).
Two agents are \emph{matched} when their current vistas are isomorphic.
They remain matched if their previous vistas, current outdegrees, and
received multisets of previous vistas agree.

In each application of \Cref{lem:two-center-graph}, \(U,W\) partition
the core and remain fixed throughout the phase, even after an agent has
been selected as a center.

\paragraph{First phase: \(\ell-1\) rounds.}
Keep the smaller core a unit bidirected star centered at \(y\).
In round \(t\) of this phase, the larger core uses the graph of \Cref{lem:two-center-graph}
with \(U\) its leaders, \(z=x'_t\), \(z'=y'\), and \(c=\ell\).
After round \(t\), all smaller leaders and \(x'_{t+1},\ldots,x'_\ell\)
share one vista, while \(y,y'_1,\ldots,y'_\ell\) share another.

By the lemma and the smaller star, the leaders being compared receive one
matching nonleader vista and have outdegree one. The nonleaders being
compared receive \(\ell\)
copies of a matching leader vista and have outdegree \(\ell\).
In the smaller star these copies come from different leaders; in the
larger graph they come from parallel arcs of \(x'_t\).
This proves the assertion by induction from the initial leader and
nonleader classes. We no longer compare \(x'_t\) after its round.

\paragraph{Second phase: \(\ell-1\) rounds.}
Apply \Cref{lem:two-center-graph} with \(c=1\) on both sides.
In round \(t\) of this phase, the smaller core has \(U\) its leaders, \(z=x_t\),
and \(z'=y\); the larger core has \(U\) its nonleaders, \(z=y'_t\),
and \(z'=x'\).
After round \(t\), \(x_{t+1},\ldots,x_\ell,x'\) share one vista, while
\(y,y'_{t+1},\ldots,y'_\ell\) share another.

The lemma gives the observations for this induction: each compared leader
receives one matching nonleader vista, and each compared nonleader receives
one matching leader vista;
all have outdegree \(\ell\). In both phases, a newly selected center
sends its matching old vista before receiving potentially different
information or learning its new outdegree. Previously selected agents
send only to this new center. Their different histories therefore do not
affect the agents still being compared.
After the two phases, \(x,x'\) match and \(y,y'\) match.

\paragraph{Adding \(g=n-\ell-1\) nonleaders.}
Add \(z_1,\ldots,z_g\) to the smaller core and
\(z'_1,\ldots,z'_g\) to the larger one.
We call these added nonleaders \emph{relays}, and each pair
\((z_i,z'_i)\), for \(1\leq i\leq g\), a \emph{relay pair}.
We describe the initial delay, the changes to the two phases, and the
final delay. For \(g=0\), omit the initial delay and leave the two phases
unchanged.

For \(g>0\), begin with \(g\) repetitions of one fixed graph on each side.
Put the nonleaders on the directed cycles
\[
 z_1\to\cdots\to z_g\to y\to z_1,
 \qquad
 z'_1\to\cdots\to z'_g\to y'_1\to\cdots\to y'_\ell\to z'_1,
\]
and join every leader bidirectionally to \(z_1\), respectively \(z'_1\).
All these arcs have multiplicity one. Each cycle has at least two agents,
so the graphs are loopless; attaching the leaders preserves strong
connectivity.

After round \(t\le g\), all original nonleaders and all
\(z_j,z'_j\) with \(j>t\) share one vista, all leaders share one vista,
and every pair \(z_j,z'_j\) is matched.
To see this, \(z_1,z'_1\) receive the same \(\ell\) leader vistas and one
matching cycle-predecessor vista, with outdegree \(\ell+1\).
Every other nonleader has outdegree one and receives its predecessor's
vista. The distinction first created at \(z_1,z'_1\) thus advances one
position per round, reaching no original nonleader before round \(g+1\).
The leaders receive the matching vistas of \(z_1,z'_1\) and have
outdegree one. This proves all three assertions by induction.

In each round of the two \((\ell-1)\)-round phases, let \(y_*\) be the nonleader
with one outgoing arc to each leader: \(y\) in the smaller
network, \(y'\) in the larger network's first phase, and
\(y'_t\) in round \(t\) of its second phase.
For \(g>0\), replace these arcs by
\[
 y_*\xrightarrow{\ell}z_1\xrightarrow{\ell}\cdots
 \xrightarrow{\ell}z_g\xrightarrow{1}\{\text{all leaders}\},
\]
using \(z'_1,\ldots,z'_g\) instead of \(z_1,\ldots,z_g\)
in the larger network. All other core arcs remain
unchanged. Every original outdegree is preserved, and every relay has
outdegree \(\ell\). Each replaced arc still has a directed path with the
same endpoints, and every relay lies on these paths, so strong
connectivity is preserved.

The phase inductions now include the assertion that every relay pair is
matched. The two agents serving as \(y_*\) have matching old vistas, even
when the larger one is the newly selected \(y'_t\). Thus corresponding
relays receive \(\ell\) copies of matching predecessor vistas and remain
matched. Each compared leader receives the last relay's matching vista
instead of \(y_*\)'s vista; the compared original nonleaders have unchanged
incoming arcs and outdegrees. The two phase invariants therefore still
hold. At their end, \(x,x'\), \(y,y'\), and every relay pair match.

\paragraph{Final cycle.}
On the smaller and larger sides, respectively, use the directed cycles
\[
\begin{aligned}
 x&\to x_1\to\cdots\to x_{\ell-1}\to y
       \to z_1\to\cdots\to z_g\to x,\\
 x'&\to x'_1\to\cdots\to x'_{\ell-1}
       \to y'_1\to\cdots\to y'_{\ell-1}\to y'
       \to z'_1\to\cdots\to z'_g\to x'.
\end{aligned}
\]
For \(g=0\), \(y,y'\) send directly to the observers.
Repeat these graphs forever. Every agent has indegree and outdegree one.
Initially the consecutive agents \(y,z_1,\ldots,z_g,x\) match their
counterparts, so a difference can advance toward the observers by only
one position per round. Their vistas remain identical for \(g+1\)
final rounds.

In fact, \(y,y'\) first differ in the first final round: their predecessors
are \(x_{\ell-1}\) and \(y'_{\ell-1}\), whose initial leader statuses differ.
This difference reaches the observers in final round \(g+2\).
The common interval therefore has exactly
\[
 g+(\ell-1)+(\ell-1)+(g+1)=2n-3
\]
rounds. In particular, the core \(g=0\) has one common final-cycle round;
the next repetition shows the first difference.

All graphs are loopless. During the two \((\ell-1)\)-round phases, arc multiplicities
are at most \(\ell\), and a selected center sends at most \(\ell\) arcs to each of at most
\(\ell\) agents, giving outdegree at most \(\ell^2\).
All other middle agents have outdegree at most \(\ell\).
The initial maximum outdegree is \(\ell+1\le\ell^2\), and the final
outdegree is one. This proves the stated structural bounds.
\end{proof}

\section{Membership in \texorpdfstring{\(\mathsf{coNP}\)}{coNP}}
\label{app:conp-membership}

We now ask whether a finite vista determines the \emph{network size},
namely the number of agents.  The
input is the abstract structure stored by an agent, which does not include its unknown
fibers.  A negative answer requires two networks of different orders
that produce the same vista.  Their orders may be exponentially large,
so an NP certificate cannot list their agents individually.  We first
bound the orders needed for such a certificate, then describe each
network by polynomially many history-tree nodes and binary anonymities.

\subsection{The \textnormal{\textsc{AtMostOneVista}} Problem}

An \emph{explicit vista} \(\mathcal V\) lists its levels
\(\mathcal L_0,\ldots,\mathcal L_t\), with one bottom node in
\(\mathcal L_t\).  The root at level \(-1\) is implicit.  Each node
at level zero has its leader or nonleader status.  For every
\(1\leq r\leq t\), the encoding gives the parent and outdegree label
of each \(v\in\mathcal L_r\), and all positive multiplicities
\(\mu_r(u,v)\) with \(u\in\mathcal L_{r-1}\).  Unlisted
multiplicities are zero; all numerical labels are encoded in binary.
The positive leader count \(\ell\) is also part of the input.

The encoding must have at most one node of each status at level zero,
distinct signatures \((d_r(v),(\mu_r(u,v))_u)\) for children of the
same parent, and exactly the nodes reached by following parents and
positive red edges backward from the bottom.  These conditions are
checkable in polynomial time. Note that these conditions describe the syntax of a vista, but do not assert that a network realizes it.

A \emph{realization} is a finite prefix of a dynamic network, with a
leader set of size \(\ell\), whose round-\(t\) vista at some agent
is isomorphic to \(\mathcal V\).  We use the isomorphisms defined in
\Cref{sec:model}.  Its order is the number of agents.  Define the
\emph{order spectrum}
\[
 \operatorname{Spec}_{\ell}(\mathcal V)
 =\{N\geq\ell:\mathcal V\text{ has a realization of order }N\}.
\]
Malformed encodings have empty spectrum.  The problem
\textsc{AtMostOneVista} asks whether this spectrum has at most one
element.  On realizable inputs, this is exactly the question whether
the agent's vista determines \(n\).

\subsection{Small Realizations}

Use the causal support \(\Gamma_0,\ldots,\Gamma_t\) from
\Cref{app:causal-support}.  In particular, \(\Gamma_0\) is the set of
agents that have influenced the observer by round \(t\).
This concerns individual agents, not whether a vista level is full: a visible node may have agents in its fiber outside
\(\Gamma_0\).

Let \(s\) be the binary length of \(\langle\mathcal V,\ell\rangle\),
let \(\nu=\sum_{r=0}^t|\mathcal L_r|\), and let \(M\) be the
largest of \(\ell\), the outdegree labels, and the red
multiplicities in \(\mathcal V\).  Put \(K=(1+\nu M)^t\).

\begin{lemma}
\label{lem:conp-bounded-orders}
If \(\mathcal V\) admits two different compatible orders, two
can be chosen at most \(K+\ell+2\), with polynomial bit length
in \(s\).
\end{lemma}

\begin{proof}
An agent represented by a node of \(\mathcal V\) at level
\(r\geq1\) constructs its vista from its own previous state,
the states received in round \(r\), and its outdegree label.
Its node has at most \(\nu\) red predecessors, each of
multiplicity at most \(M\).  Unfolding the previous and received
states backward for \(t\) rounds therefore gives
\(|\Gamma_0|\leq(1+\nu M)^t=K\).
If two different compatible orders have realizations with
\(\Gamma_0=V\), both orders are at most \(K\).

Otherwise, take a realization with \(\Gamma_0\neq V\).
Retain all agents in \(\Gamma_0\) and all leaders; merge the
remaining nonleaders, if any, into a single nonleader.
Replace each arc by the arc between the images of its endpoints,
retaining every parallel copy.  Contraction preserves strong
connectivity and every retained agent's outdegree.
No merged agent sends into \(\Gamma_r\) in round \(r\),
since such a sender would belong to
\(\Gamma_{r-1}\subseteq\Gamma_0\).
Thus all states needed for \(\mathcal V\) remain unchanged.
The resulting realization has at most \(K+\ell+1\) agents
and still has an agent outside the causal support.
Loops created by contraction are allowed.

Now add one nonleader.  In each round \(r\), strong connectivity
gives an arc from the nonempty proper set \(\Gamma_r\) to its
complement.  Replace one copy by a two-arc path through the new
agent.  This preserves the original tail's outdegree and changes
no incoming arc at any agent in \(\Gamma_r\).
Induction through the rounds preserves every state needed for
\(\mathcal V\), and the new agent remains outside the causal
support.  We have obtained two different compatible orders,
both at most \(K+\ell+2\).

Finally, \(t,\nu\), and \(\log M\) are bounded by
\(s\), so \(K+\ell+2\) has polynomial bit length in \(s\).
\end{proof}

\begin{corollary}
\label{cor:conp-pumping}
Let $\mathcal V$ be a vista with $\ell$ leaders. If it has a
realization of order $N$ whose observer has a proper initial causal
support $\Gamma_0\ne V$, then
\[
  \{N,N+1,N+2,\ldots\}\subseteq
  \operatorname{Spec}_{\ell}(\mathcal V).
\]
Moreover, with $K=(1+\nu M)^t$ as above, either all compatible orders
are at most $K$, or there is an integer $N_*\le K+\ell+1$ such that
\[
  \{N_*,N_*+1,N_*+2,\ldots\}\subseteq
  \operatorname{Spec}_{\ell}(\mathcal V).
\]
In particular, every order spectrum is finite or cofinite.
\end{corollary}

\begin{proof}
For the first assertion, repeat the arc-subdivision construction in
the proof of \Cref{lem:conp-bounded-orders}. Each added nonleader
leaves the observer's vista unchanged and remains outside its
initial causal support. Thus every order at least \(N\) is compatible.

If no realization has a proper initial causal support, every
compatible order is at most \(K\). Otherwise, the contraction
construction in the same proof gives a realization of order
\(N_*\leq K+\ell+1\) with proper initial causal support.
The first assertion then makes every order at least \(N_*\)
compatible. Hence the spectrum is finite in the first case and
cofinite in the second.
\end{proof}

\subsection{Constructing the Network}

A certificate will describe a full history-tree prefix, including
nodes absent from the input vista.  It gives each node a positive
binary anonymity and specifies its parent, outdegree, and incoming
multiplicities.  To verify that these tables come from strongly
connected round graphs, we need the following elementary construction.

\begin{lemma}
\label{lem:conp-balanced-vertex}
Let \(D\) be a directed multigraph on \(V\cup\{z\}\), with
\(V\neq\varnothing\).  Suppose every ordered pair of vertices in
\(V\) is joined by a directed path, \(z\) has no loop, and
\(d^-(z)=d^+(z)\).  The arcs at \(z\) can be paired so that replacing
each pair \(a\to z\to b\) by \(a\to b\) leaves a strongly
connected graph on \(V\).
\end{lemma}

\begin{proof}
Pair the arcs arbitrarily and call the resulting graph \(G\).
If it is not strongly connected, choose distinct strongly connected
components \(A,B\), with no arc entering \(A\) and no arc leaving
\(B\).  Reachability in \(D\) forces an arc from \(z\) into
\(A\).  Its paired shortcut \(a\to b\) lies wholly in \(A\),
since no arc of \(G\) enters \(A\).  Similarly, an arc from
\(B\) to \(z\) gives a paired shortcut \(c\to d\) wholly in \(B\).

Switch these pairings, replacing the shortcuts by \(a\to d\)
and \(c\to b\).  Simple paths \(b\leadsto a\) within \(A\)
and \(d\leadsto c\) within \(B\) avoid the removed shortcuts.
The removed arcs therefore have replacement paths
\[
 a\to d\leadsto c\to b,
 \qquad c\to b\leadsto a\to d.
\]
No old reachability is lost, and \(A,B\) merge.  Repeat until only
one component remains.  A shortcut that is a loop uses an empty
internal path in this argument.
\end{proof}

\begin{lemma}
\label{lem:conp-transition}
Let \(\mathcal T_{r-1},\mathcal T_r\) be the old and new node sets
of a proposed full history-tree prefix, with parent map
\(\pi:\mathcal T_r\to\mathcal T_{r-1}\).  New node \(v\) has
anonymity \(\anon(v)\geq1\) and outdegree \(d(v)\).  Each agent in
its fiber must receive \(\mu(u,v)\) arcs from the fiber of old
node \(u\).  Assume the flow equations
\begin{equation}
 \sum_{v:\pi(v)=u}\anon(v)d(v)
 =\sum_{w\in\mathcal T_r}\anon(w)\mu(u,w)
 \qquad(u\in\mathcal T_{r-1}).
 \label{eq:conp-full-flow}
\end{equation}
Put \(N=\sum_v \anon(v)\).  For \(N>1\), these data have a strongly
connected round realization exactly when every new node has positive
outdegree and positive incoming degree, and the digraph on
\(\mathcal T_r\) with arcs
\(v\to w\) whenever \(d(v)>0\) and \(\mu(\pi(v),w)>0\)
is strongly connected.  For \(N=1\), balance alone suffices.
\end{lemma}

\begin{proof}
Necessity follows by projecting physical paths onto their nodes.
For sufficiency, create the \(N\) agents and an auxiliary vertex
\(z_u\) for each old node \(u\) with positive flow total.
An agent in the fiber of \(v\) has \(d(v)\) arcs to
\(z_{\pi(v)}\); vertex \(z_u\) has \(\mu(u,w)\) arcs to
each agent in the fiber of \(w\).  Equation
\eqref{eq:conp-full-flow} makes each auxiliary vertex balanced.
Every arc of the stated digraph lifts to a two-arc path between
any agents in the corresponding fibers.  Positive degrees and
strong connectivity of that digraph make the auxiliary graph
strongly connected.

Apply \Cref{lem:conp-balanced-vertex} to remove the auxiliary
vertices one at a time.  There are no arcs between them, so removing
one leaves all others balanced.  Each replacement preserves its
source's outdegree and its target's receipt from the specified old
fiber.  Loops and parallel arcs are permitted.  For \(N=1\),
pair the single agent's outgoing arcs and incoming requirements
directly as loops.
\end{proof}

The flow equation in the previous lemma is exactly \eqref{eq:flow},
with both visible and hidden nodes included. The verifier checks the
lemma's conditions directly on the binary-encoded node tables, without
constructing the potentially exponentially large network.

\subsection{Compressing the History Tree}

We must still show that any realization of \(\mathcal V\) can be
replaced by one of the same order whose full history-tree prefix
through level \(t\) has only polynomially many nodes.
We do this by rewiring arcs ending at agents in hidden fibers,
while preserving all visible fibers and observations and keeping
every round graph strongly connected.

\begin{lemma}
\label{lem:conp-normal-form}
If \(\mathcal V\) has a realization of order \(N\), it has one
of the same order and leader count whose full levels
\(\mathcal T_0,\ldots,\mathcal T_t\) satisfy
\[
 |\mathcal T_r|\leq\sum_{j=0}^r|\mathcal L_j|+r+2.
\]
Its anonymities, outdegrees, and multiplicities have
\(O(s+\log N)\) bits.  The entire full-prefix description has
length polynomial in \(s+\log N\).
\end{lemma}

\begin{proof}
Fix a realization of \(\mathcal V\). We keep the same agents and
leader statuses, but rebuild the round graphs one round at a time.
At each level, we partition the agents into sets called
\emph{blocks}, which the construction will make into the fibers
of the new history-tree nodes. Blocks may split from one level
to the next, but never merge. Every visible fiber remains an
unchanged block; the other blocks will be fibers of hidden nodes.
At level zero, use one block for all leaders and, if nonempty,
one for all nonleaders.

Suppose the partition at level \(r-1\) has been constructed.
Within each block \(C\), retain each original visible fiber at
level \(r\) as a separate block, and put all remaining agents
of \(C\) into one block \(H_C\), omitting it if empty.
A hidden block at level \(r-1\) contains no visible fiber at
level \(r\), since every visible node has a visible parent.
If there are no hidden agents at level \(r\), keep the original
round graph: all blocks at level \(r-1\) are then unchanged
visible fibers.

Otherwise, strong connectivity gives an original round-\(r\) arc
from an agent \(q\) outside the visible fibers at level \(r\)
to an agent inside them. The original node at level \(r-1\)
whose fiber contains \(q\) is therefore visible.
Since visible fibers are preserved, this fiber is also a block
\(P\) of the new partition at level \(r-1\).
Split \(H_P\) into \(\{q\}\) and \(H_P\setminus\{q\}\),
omitting the latter if empty.

Keep all original arcs ending at visible new agents.  No agent
in a hidden old block has such an arc, since its old node would
otherwise be visible.  For each ordered pair supported by an
original arc with a hidden target, retain one arc from its source
but redirect it to \(q\).  For a source with visible new node
\(v\), these retained arcs are a subset of its original outgoing
arcs.  Add arcs to \(q\) until its outdegree again equals the
displayed value \(d_r(v)\).

Choose \(D\) larger than all currently retained degrees and all
displayed new degrees.  Give every hidden agent other than \(q\)
outdegree \(D\), by adding arcs to \(q\).  Add one arc from
\(q\) to each other hidden agent, then pad \(q\)'s outdegree
with loops to a value \(D_q>D\) exceeding its current degree.
Every block \(H_C\) other than \(\{q\}\) now has one uniform
signature: its agents have degree \(D\), receive one state from
old block \(P\), and receive nothing else.  These degrees
distinguish hidden children from visible siblings and distinguish
\(q\) from its hidden siblings.  Visible children retain their
incoming multiplicities and displayed degrees.  Thus the new
partition is exactly the partition into histories for this round,
and the observer's vista is unchanged.

The round remains strongly connected.  Every hidden agent other
than \(q\) has an arc to \(q\) and one from \(q\).  To reach
\(q\) from a visible agent, follow an original path until its
first hidden target; that arc was redirected to \(q\).  To reach
a visible agent from \(q\), take an original path and its last
hidden vertex \(z\).  The remaining visible suffix was retained;
if \(z\neq q\), prepend the added arc \(q\to z\).
All visible incoming records are preserved, and no hidden old
block sends to a visible new block, so no additional predecessor
enters the vista.  This proves the induction invariant.

The new partition has at most \(|\mathcal L_r|\) visible blocks,
one hidden block per old block, and the extra singleton \(q\).
Starting with at most two blocks gives the claimed node bound.
For bit lengths, all retained arcs to visible targets total at most
\(N\nu M\) per source, and at most \(N\) distinct hidden-target
arcs per source are retained.  Choose \(D,D_q\) by adding one to
the required maxima.  All new degrees and incoming multiplicities
are polynomially bounded in \(N,\nu,M\), and hence have
\(O(s+\log N)\) bits.  Since every vista level is nonempty,
\(t+1\leq\nu\); the tables have \(O(\nu^2)\) nodes and
\(O(\nu^3)\) multiplicity entries.
\end{proof}

\begin{lemma}
\label{lem:conp-fixed-order}
Whether \(N\in\operatorname{Spec}_{\ell}(\mathcal V)\) is in
\(\mathsf{NP}\), with certificate length polynomial in
\(s+\log N\).
\end{lemma}

\begin{proof}
Guess the full-prefix tables of \Cref{lem:conp-normal-form},
a node \(b\) at level \(t\) to serve as the vista's bottom node,
and an identification of the input vista with the backward closure
of \(b\). Check positive integer anonymities, total anonymity \(N\)
at each level, and exactly \(\ell\) leaders. Check disjoint refinement: each parent's
anonymity must equal the sum of its children's anonymities.
At level zero, check that each occurring status has exactly one node.
At later levels, check that each node has the same leader status
as its parent (H2), and check nonnegative degrees and multiplicities,
the full flow equations \eqref{eq:conp-full-flow}, distinct sibling
signatures, and the realizability conditions of
\Cref{lem:conp-transition}. Finally, check that the supplied
identification preserves all edge types, labels, and statuses
and maps exactly the backward closure of \(b\) onto the input vista.

All checks take time polynomial in the binary table length.
Disjoint refinement allows us to partition the same
set of \(N\) agents into fibers at successive levels, dividing
each parent's fiber among its children.
\Cref{lem:conp-transition} realizes every round on those agents,
while distinct sibling signatures ensure that the resulting
history tree has exactly the proposed nodes.
Thus every accepted certificate gives a realization of
\(\mathcal V\). Conversely, \Cref{lem:conp-normal-form}
supplies a certificate for every compatible order.
\end{proof}

\subsection{Proof of Membership}

\begin{theorem}
\label{thm:conp-vista-membership}
In the model of \Cref{sec:model}, \textnormal{\textsc{AtMostOneVista}}
belongs to \(\mathsf{coNP}\).
\end{theorem}

\begin{proof}
For the complementary problem, guess two different compatible
orders and their certificates from \Cref{lem:conp-fixed-order}.
By \Cref{lem:conp-bounded-orders}, the orders can be chosen
at most \(K+\ell+2\), with polynomial bit length in \(s\).
Both certificates therefore have polynomial length.
Reject malformed inputs, whose spectra are empty.
Taking complements proves the theorem.
\end{proof}

There is also a network-level version.  An input explicitly lists
the agents, leader set, and round graphs through round \(t\), with
binary arc multiplicities. The problem
\textsc{SomeAgentKnows} asks whether at least one round-\(t\)
vista determines the network size.

\begin{theorem}
\label{thm:conp-network-membership}
\textnormal{\textsc{SomeAgentKnows}} belongs to \(\mathsf{coNP}\).
\end{theorem}

\begin{proof}
Construct all round-\(t\) vistas by computing the successive
partitions and history-tree labels.  This takes polynomial time
in the explicitly listed network prefix.  Its order \(n\) is
compatible with every one of these vistas.

If no agent's vista determines \(n\), each vista has at least
two compatible orders.  By \Cref{lem:conp-bounded-orders}, two
can be chosen with polynomial bit length, and at least one
differs from \(n\).  Guess one such alternative order and its
fixed-order certificate for every listed agent.
The constructed vistas have polynomial length in the network
input, so the whole list has polynomial length.
Verify the certificates using \Cref{lem:conp-fixed-order}.
\end{proof}

\section{\texorpdfstring{\(\mathsf{coNP}\)}{coNP}-Hardness}
\label{app:conp-hardness}

We prove hardness for \textsc{AtMostOneVista}, defined in
\Cref{sec:network-wise-complexity}, and \textsc{SomeAgentKnows},
defined in Appendix~\ref{app:conp-membership}.
We do not require a second compatible realization to have smaller order.
The construction uses binary-encoded parallel arcs, as allowed by the
input model, and produces loopless, strongly connected round graphs.

\subsection{Encoding Exact Cover}

Reduce from the complement of Exact Cover by 3-Sets
(\textsc{X3C})~\cite[Problem~SP2]{GJ79}.
Let \(U=\{0,\ldots,3q-1\}\), and let \(S_1,\ldots,S_m\) be distinct
three-element subsets of \(U\).
We may assume \(q\geq2\): duplicate triples can be deleted, and adding
three fresh elements with their unique triple increases \(q\) by one
without changing whether an exact cover exists.

Set
\[
 B=q+3,\qquad K=B^{3q},\qquad
 W=\sum_{e=0}^{3q-1}B^e,\qquad A=qK+W,
\]
and define
\[
 c_0=A,\qquad
 c_j=K+\sum_{e\in S_j}B^e\quad(1\leq j\leq m).
\]
We have \(0<W<K\), all \(c_i>K\), and \(A<(q+1)K\).
The \(c_i\) are distinct: the triples have distinct base-\(B\) digit
vectors, and
\[
 c_j<2K\leq qK<A=c_0 \qquad (j\geq1).
\]

\begin{lemma}
\label{lem:conp-digits}
For nonnegative integers \(x_0,\ldots,x_m\), the equation
\begin{equation}
 A x_0+\sum_{j=1}^m c_jx_j=A
 \label{eq:target}
\end{equation}
has the reference solution \(x_0=1,\ x_1=\cdots=x_m=0\).
Every other solution has \(x_0=0\), and its nonzero coordinates
\(x_j\) select an exact cover. Conversely, every exact cover gives
such a solution. In every solution,
\(\sum_{i=0}^m x_i\leq q\).
\end{lemma}

\begin{proof}
Since every coefficient is at least \(K\) and \(A<(q+1)K\),
the sum of the coordinates is at most \(q\).
Equation~\eqref{eq:target} gives \(x_0\leq1\); if \(x_0=1\),
positivity of the other coefficients gives the reference solution.

If \(x_0=0\), write
\[
 k=\sum_{j=1}^m x_j,\qquad
 d_e=\sum_{j:e\in S_j}x_j.
\]
Then
\[
 kK+\sum_{e=0}^{3q-1}d_e B^e
   =qK+\sum_{e=0}^{3q-1}B^e.
\]
Here \(0\leq d_e\leq k\leq q<B\), so no carries occur in base \(B\).
The digit at position \(3q\) gives \(k=q\), and the remaining digits
give \(d_e=1\) for every \(e\). Each \(x_j\) is consequently zero or
one, and the selected triples form an exact cover. The converse
follows by summing the coefficients of a cover.
\end{proof}

\subsection{The Reference Network}

There is one leader \(\lambda\) and groups \(0,\ldots,m\) of nonleaders. Write
\[
 R=1+(m+1)B,\qquad N_0=R+1,\qquad
 C=B\sum_{i=0}^m c_i+A.
\]
The reference group sizes are
\[
 a_0^0=B+1,\qquad a_j^0=B\quad(1\leq j\leq m).
\]
Thus the reference network has \(N_0\) agents and
\(\sum_i c_i a_i^0=C\).

More generally, for group sizes \(a_i\geq B\) with
\(\sum_i c_i a_i=C\), list group \(i\) as
\(x_{i,1},\ldots,x_{i,a_i}\).
The \emph{path graph} has the arcs
\[
 x_{i,1}\to x_{i,2}\to\cdots\to x_{i,a_i}\to\lambda,
\]
one copy of each, together with \(c_i\) parallel arcs
\(\lambda\to x_{i,k}\) for every group member.
Use this same graph for rounds \(1,\ldots,B-1\).
Every nonleader has outdegree one, and the leader has outdegree \(C\).
The graph is loopless and strongly connected: the leader reaches every
agent directly, and every agent has a path back to the leader.

In round \(B\) of the reference network, replace the path arcs by one arc from
every nonleader to the leader. Keep the same leader-to-agent arcs.
Call this the \emph{ordinary star}.
Let \(\mathcal G^0\) be this \(B\)-round reference prefix, and let
\(\mathcal V\) be its final leader vista. These are the outputs for the
network and vista reductions, respectively.

\begin{lemma}
\label{lem:conp-path-prefix}
For any sizes \(a_i\geq B\) with \(\sum_i c_i a_i=C\), the labeled
history-tree prefix through round \(B-1\), without fiber
cardinalities, is the same.
After round \(r\), for \(1\leq r\leq B-1\), group \(i\) has
the \(r\) singleton classes
\[
 \{x_{i,1}\},\ldots,\{x_{i,r}\}
\]
and the single nonempty suffix class
\(\{x_{i,r+1},\ldots,x_{i,a_i}\}\).
All groups have distinct histories.
\end{lemma}

\begin{proof}
In round one, the first path member receives only \(c_i\) leader
states; every other member also receives one nonleader state.
The leader receives \(m+1\) copies of the initial nonleader state.
Its outdegree is \(C\). Hence all these records are independent of the
group sizes, and the distinct \(c_i\) distinguish the groups.

Inductively, previously distinct classes cannot merge, since the
previous history is retained. At round \(r+1\), the first member of
the old suffix receives the previous state of the singleton at
position \(r\), while every remaining suffix member receives the old
suffix state. This separates one further position. Every member
still receives \(c_i\) copies of the same previous leader state and
has outdegree one.
The last member of each path is in its old suffix, so the leader
receives exactly one old suffix state from each group and still has
outdegree \(C\). These records are independent of the lengths.
This proves both the partition and the equality of its labels.
\end{proof}

In particular, at level \(B-1\) there are exactly \(B\) nonleader
nodes per group and one leader node, for a total of \(R\) nodes.
The reference network's ordinary star makes every one of those nodes visible
in \(\mathcal V\): every nonleader sends to the leader, so every
nonleader node is a red-edge predecessor of the new leader node.
The old leader node is its parent. All earlier nodes are visible by
following parents.

\paragraph{Polynomial size.}
The reference network has \(O((m+1)q)\) agents and \(O(q)\) rounds, with
\(O((m+1)q^2)\) nonzero arc entries over all round graphs.
The numbers \(c_i\) have \(O(q\log(q+3))\) bits, and \(C\) has
\(O(q\log(q+3)+\log(m+1))\) bits.
The vista has \(O((m+1)q^2)\) nodes and edges.
It can be computed by successive history partitions, using binary
multiplicities throughout. Both outputs are therefore computable in
polynomial time; parallel arcs are not expanded into individual
input entries.

\subsection{The Possible Network Sizes}

Fix any one-leader realization of \(\mathcal V\), on any agent set
\(V\). We do not assume that its groups are paths or that nodes
with singleton fibers in the reference network still have singleton fibers
in this realization.

For each \(1\leq r\leq B-1\), let \(S_r\subseteq V\) be the union
of the fibers of the nodes visible at level \(r\) of \(\mathcal V\).
Parents are visible, so \(S_{r+1}\subseteq S_r\).
The unique leader belongs to every \(S_r\).

At level one, each group is represented by \emph{two} nodes, its
first-position node and its suffix node. Let \(a_i\) be the sum of
the cardinalities of these two fibers in the arbitrary realization.
Their visible descendants at level \(B-1\) include \(B\) disjoint
nonempty fibers. Therefore
\[
 a_i\geq B.
\]
Both first-level nodes of group \(i\) prescribe \(c_i\) incoming
copies of the unique initial leader state. The visible leader's
first-round outdegree is \(C\), so
\[
 \sum_{i=0}^m c_i a_i\leq C.
\]
This is only an inequality until we prove that level one is full.
Writing \(x_i=a_i-B\geq0\) gives
\[
 \sum_{i=0}^m c_i x_i\leq A,
 \qquad
 \sum_{i=0}^m x_i\leq q,
\]
where the second bound again uses \(c_i\geq K\) and
\(A<(q+1)K\). Hence
\begin{equation}
 |S_1|\leq R+q,\qquad |S_{B-1}|\geq R.
 \label{eq:visible-bounds}
\end{equation}

\begin{lemma}
\label{lem:conp-first-full}
In every realization of \(\mathcal V\), \(S_1=V\).
\end{lemma}

\begin{proof}
Suppose \(S_1\) is proper. Every \(S_r\subseteq S_1\) is then
proper and nonempty. If \(S_{r+1}=S_r\), strong connectivity of
round \(r+1\) gives an arc from \(V\setminus S_r\) into this set.
Its head lies in a visible level-\((r+1)\) fiber, so its tail's
level-\(r\) history must be a visible incoming predecessor.
The tail would belong to \(S_r\), a contradiction.
Thus every inclusion \(S_{r+1}\subseteq S_r\) is strict.

There are \(B-2=q+1\) such transitions between levels \(1\)
and \(B-1\). Consequently~\eqref{eq:visible-bounds} gives
\[
 R\leq |S_{B-1}|
   \leq |S_1|-(q+1)
   \leq (R+q)-(q+1)=R-1,
\]
a contradiction.
\end{proof}

Since level one is full, the first-round leader outdegree is
now accounted for exactly by the displayed nonleader fibers; the
leader's displayed incoming record contains no leader state, so there
is no leader loop. Thus
\[
 \sum_{i=0}^m c_i a_i=C,
 \qquad\text{equivalently}\qquad
 A x_0+\sum_{j=1}^m c_jx_j=A.
\]
\Cref{lem:conp-digits} applies. If there is no exact cover, it forces
\(x_0=1\) and all other coordinates zero, so every realization has
order \(N_0=R+1\). If any other solution occurs, the source
instance has an exact cover, and its order is \(R+q\).

Combining this converse with the constructions below will give
\[
 \operatorname{Spec}_1(\mathcal V)=
 \begin{cases}
  \{N_0\},&\text{if the instance has no exact cover},\\
  \{N_0,N_0+q-1\},&\text{if the instance has an exact cover}.
 \end{cases}
\]
This argument shows that level one is full. It does not require later
visible levels to be full, or the actual order to be known in advance.

\subsection{The Alternative Networks}

Suppose the source instance has an exact cover \(J\subseteq
\{1,\ldots,m\}\), with \(|J|=q\). Define alternative group sizes by
\[
 a_0'=B,\qquad
 a_j'=B+\mathbf 1_{\{j\in J\}}\quad(j\geq1).
\]
Their order is
\[
 N_1=R+q=N_0+q-1\ne N_0.
\]
Moreover,
\[
 \sum_i c_i a_i'
   =B\sum_i c_i+\sum_{j\in J}c_j
   =B\sum_i c_i+A=C.
\]
By \Cref{lem:conp-path-prefix}, these sizes realize exactly the same
labeled path prefix through round \(B-1\).

At its last level, every reference nonleader class is a singleton
except group \(0\)'s suffix, which has two members.
In the alternative prefix, group \(0\)'s suffix is a singleton,
each selected group's suffix has two members, and every other
class is a singleton.

\paragraph{The leader's vista.}
Keep the usual \(c_i\) leader-to-agent arcs in the alternative
final round, and set the nonleader arcs as follows.
The single member of group \(0\)'s suffix sends two parallel arcs
to the leader. Within each selected group's two-member suffix,
choose one member to send to the other, and let the other send
one arc to the leader. Every remaining nonleader sends one arc
to the leader.

The leader receives two copies of group \(0\)'s old suffix state,
one copy of each selected group's old suffix state, and one copy
of every other old nonleader state. These are exactly its receipts
in the reference network's ordinary star. Its previous history and current
outdegree \(C\) also agree. Thus its final vista is \(\mathcal V\).

This alternative graph is loopless and strongly connected.
The leader reaches every agent directly, and every agent reaches
it in at most two arcs. The doubled return arc changes that
sender's current outdegree, but not the previous history it sends;
this is exactly where late outdegree awareness is used.

\paragraph{The nonleaders' vistas.}
For this purpose, instead use the ordinary star on the alternative
prefix. Every reference nonleader has a counterpart in the same
old history class, because every class is nonempty in both
prefixes. Such a counterpart has the same previous history,
current outdegree one, and incoming multiset consisting of
\(c_i\) copies of the previous leader history.
Their final vistas therefore agree.

A single ordinary-star alternative reproduces every reference
nonleader vista simultaneously. Its leader need not reproduce the
reference leader; the preceding alternative final graph does that.
No one-to-one correspondence between the two agent sets is required.

\begin{theorem}
\label{thm:conp-vista-complete}
\textnormal{\textsc{AtMostOneVista}} is \(\mathsf{coNP}\)-complete, even on realizable vistas with one leader.
\end{theorem}

\begin{proof}
Output \(\mathcal V\). It always has the reference realization of
order \(N_0\). If the source instance has no exact cover,
\Cref{lem:conp-first-full} and the exact integer equation show that
every realization has order \(N_0\).
If it has an exact cover, the adjusted alternative final round
above gives a realization of order \(N_1\ne N_0\).
Thus the complement of \textsc{X3C} reduces to \textsc{AtMostOneVista}.
Membership follows from \Cref{thm:conp-vista-membership}.
\end{proof}

\begin{theorem}
\label{thm:conp-network-complete}
\textnormal{\textsc{SomeAgentKnows}} is \(\mathsf{coNP}\)-complete, even for one-leader inputs whose round graphs
are loopless.
\end{theorem}

\begin{proof}
Output the explicit reference prefix \(\mathcal G^0\), with its
ordinary-star final round. If the source instance has no exact
cover, the leader has only the compatible order \(N_0\), so the
output is a yes-instance.

If an exact cover exists, the adjusted alternative final round
gives the reference leader a compatible order \(N_1\), while the
ordinary-star alternative gives every reference nonleader the
same compatible order \(N_1\).
Since all their vistas are also realized by the reference order
\(N_0\ne N_1\), no reference agent knows the order.
This reduces the complement of \textsc{X3C} to \textsc{SomeAgentKnows}.
Membership follows from \Cref{thm:conp-network-membership}.
\end{proof}

Both reductions use loopless, strongly connected round graphs.
The leader-to-agent multiplicities may be exponentially large,
but have polynomial binary length.

\bibliographystyle{plainurl}
\bibliography{references}

\end{document}